\pdfoutput=1
\documentclass[12pt,a4paper]{article}

\usepackage{ifthen} 
\newboolean{pdflatex}
\setboolean{pdflatex}{true} 

\newboolean{articletitles}
\setboolean{articletitles}{true} 

\newboolean{uprightparticles}
\setboolean{uprightparticles}{false} 

\def\paperauthors{LHCb collaboration} 
\def\paperasciititle{Search for rare decays of D0 mesons into two muon} 
\def\papertitle{Search for rare decays of \Dz mesons into two muons} 
\def\paperkeywords{{High Energy Physics}, {LHCb}} 
\def\papercopyright{\the\year\ CERN for the benefit of the LHCb collaboration} 
\def\paperlicence{CC BY 4.0 licence}
\def\paperlicenceurl{https://creativecommons.org/licenses/by/4.0/}

\usepackage[top=1in, bottom=1.25in, left=1in, right=1in]{geometry}

\usepackage{microtype}
\usepackage{lineno}  
\usepackage{xspace} 
\usepackage{caption} 

\usepackage{graphicx}  
\usepackage{color}
\usepackage{colortbl}
\graphicspath{{./figs/}} 

\usepackage{amsmath} 
\usepackage{amssymb}
\usepackage{amsfonts}
\usepackage{upgreek} 

\newcommand*\patchAmsMathEnvironmentForLineno[1]{%
\expandafter\let\csname old#1\expandafter\endcsname\csname #1\endcsname
\expandafter\let\csname oldend#1\expandafter\endcsname\csname
end#1\endcsname
 \renewenvironment{#1}%
   {\linenomath\csname old#1\endcsname}%
   {\csname oldend#1\endcsname\endlinenomath}%
}
\newcommand*\patchBothAmsMathEnvironmentsForLineno[1]{%
  \patchAmsMathEnvironmentForLineno{#1}%
  \patchAmsMathEnvironmentForLineno{#1*}%
}
\AtBeginDocument{%
\patchBothAmsMathEnvironmentsForLineno{equation}%
\patchBothAmsMathEnvironmentsForLineno{align}%
\patchBothAmsMathEnvironmentsForLineno{flalign}%
\patchBothAmsMathEnvironmentsForLineno{alignat}%
\patchBothAmsMathEnvironmentsForLineno{gather}%
\patchBothAmsMathEnvironmentsForLineno{multline}%
\patchBothAmsMathEnvironmentsForLineno{eqnarray}%
}

\usepackage{hyperxmp}

\usepackage[pdftex,
            pdfauthor={\paperauthors},
            pdftitle={\paperasciititle},
            pdfkeywords={\paperkeywords},
            pdfcopyright={Copyright (C) \papercopyright},
            pdflicenseurl={\paperlicenceurl}]{hyperref}

\usepackage[colorinlistoftodos,textsize=scriptsize]{todonotes}

\usepackage[bottom,flushmargin,hang,multiple]{footmisc}

\usepackage[all]{hypcap} 

\usepackage{xspace} 
\usepackage{upgreek}

\def\lhcb   {\mbox{LHCb}\xspace}

\def\MagUp {\mbox{\em Mag\kern -0.05em Up}\xspace}

\ifthenelse{\boolean{uprightparticles}}%
{

 \def\Pmu         {\ensuremath{\upmu}\xspace}

 \def\Ppi         {\ensuremath{\uppi}\xspace}

 \def\Ppsi        {\ensuremath{\uppsi}\xspace}

 \def\PDelta      {\ensuremath{\Delta}\xspace}                 
 \def\PXi         {\ensuremath{\Xi}\xspace}                 
 \def\PLambda     {\ensuremath{\Lambda}\xspace}                 
 \def\PSigma      {\ensuremath{\Sigma}\xspace}                 
 \def\POmega      {\ensuremath{\Omega}\xspace}                 
 \def\PUpsilon    {\ensuremath{\Upsilon}\xspace}

 \def\PB      {\ensuremath{\mathrm{B}}\xspace}                 
                  
 \def\PD      {\ensuremath{\mathrm{D}}\xspace}

 \def\PJ      {\ensuremath{\mathrm{J}}\xspace}                 
 \def\PK      {\ensuremath{\mathrm{K}}\xspace}

 \def\Pi      {\ensuremath{\mathrm{i}}\xspace}

 \def\Pp      {\ensuremath{\mathrm{p}}\xspace}

 \def\Ps      {\ensuremath{\mathrm{s}}\xspace}

 \def\thebaroffset{0.0em}
}
{

 \def\Pmu         {\ensuremath{\mu}\xspace}

 \def\Ppi         {\ensuremath{\pi}\xspace}

 \def\Ppsi        {\ensuremath{\psi}\xspace}                 
                  
 \mathchardef\PDelta="7101
 \mathchardef\PXi="7104
 \mathchardef\PLambda="7103
 \mathchardef\PSigma="7106
 \mathchardef\POmega="710A
 \mathchardef\PUpsilon="7107
                  
 \def\PB      {\ensuremath{B}\xspace}                 
                  
 \def\PD      {\ensuremath{D}\xspace}

 \def\PJ      {\ensuremath{J}\xspace}                 
 \def\PK      {\ensuremath{K}\xspace}

 \def\Pi      {\ensuremath{i}\xspace}

 \def\Pp      {\ensuremath{p}\xspace}

 \def\Ps      {\ensuremath{s}\xspace}

 \def\thebaroffset{0.18em}
}
\newcommand{\offsetoverline}[2][\thebaroffset]{\kern #1\overline{\kern -#1 #2}}%

\makeatletter
\ifcase \@ptsize \relax
  \newcommand{\miniscule}{\@setfontsize\miniscule{4}{5}}
\or
  \newcommand{\miniscule}{\@setfontsize\miniscule{5}{6}}
\or
  \newcommand{\miniscule}{\@setfontsize\miniscule{5}{6}}
\fi
\makeatother

\DeclareRobustCommand{\optbar}[1]{\shortstack{{\miniscule (\rule[.5ex]{1.25em}{.18mm})}
  \\ [-.7ex] $#1$}}

\def\mup        {{\ensuremath{\Pmu^+}}\xspace}
\def\mun        {{\ensuremath{\Pmu^-}}\xspace} 

\def\mumu       {{\ensuremath{\Pmu^+\Pmu^-}}\xspace}

\def\squark    {{\ensuremath{\Ps}}\xspace}

\def\pion   {{\ensuremath{\Ppi}}\xspace}

\def\pip    {{\ensuremath{\pion^+}}\xspace}
\def\pim    {{\ensuremath{\pion^-}}\xspace}

\def\kaon    {{\ensuremath{\PK}}\xspace}

\def\KorKbar {\kern \thebaroffset\optbar{\kern -\thebaroffset \PK}{}\xspace}

\def\Kp      {{\ensuremath{\kaon^+}}\xspace}
\def\Km      {{\ensuremath{\kaon^-}}\xspace}

\def\Dbar    {{\ensuremath{\offsetoverline{\PD}}}\xspace}
\def\D       {{\ensuremath{\PD}}\xspace}

\def\DorDbar {\kern \thebaroffset\optbar{\kern -\thebaroffset \PD}\xspace}
\def\Dz      {{\ensuremath{\D^0}}\xspace}
\def\Dzb     {{\ensuremath{\Dbar{}^0}}\xspace}
\def\Dp      {{\ensuremath{\D^+}}\xspace}
\def\Dm      {{\ensuremath{\D^-}}\xspace}

\def\DpDm    {\ensuremath{\Dp {\kern -0.16em \Dm}}\xspace}
\def\Dstar   {{\ensuremath{\D^*}}\xspace}

\def\Dstarp  {{\ensuremath{\D^{*+}}}\xspace}

\def\Ds      {{\ensuremath{\D^+_\squark}}\xspace}

\def\B       {{\ensuremath{\PB}}\xspace}

\def\BorBbar {\kern \thebaroffset\optbar{\kern -\thebaroffset \PB}\xspace}

\def\Bd      {{\ensuremath{\B^0}}\xspace}

\def\BdorBdbar {\kern \thebaroffset\optbar{\kern -\thebaroffset \Bd}\xspace}
\def\Bu      {{\ensuremath{\B^+}}\xspace}

\def\Bs      {{\ensuremath{\B^0_\squark}}\xspace}

\def\BsorBsbar {\kern \thebaroffset\optbar{\kern -\thebaroffset \Bs}\xspace}

\def\jpsi     {{\ensuremath{{\PJ\mskip -3mu/\mskip -2mu\Ppsi}}}\xspace}

\def\Y#1S{\ensuremath{\PUpsilon{(#1S)}}\xspace}

\def\proton      {{\ensuremath{\Pp}}\xspace}

\def\LorLbar     {\kern \thebaroffset\optbar{\kern -\thebaroffset \PLambda}\xspace}

\newcommand{\decay}[2]{\ensuremath{#1\!\to #2}\xspace} 

\def\to                 {\ensuremath{\rightarrow}\xspace}

\newcommand{\dm}{{\ensuremath{\Delta m}}\xspace}

\def\AT#1     {\ensuremath{A_{\mathrm{T}}^{#1}}\xspace}           

\def\C#1      {\ensuremath{\mathcal{C}_{#1}}\xspace}                       
\def\Cp#1     {\ensuremath{\mathcal{C}_{#1}^{'}}\xspace}                    
\def\Ceff#1   {\ensuremath{\mathcal{C}_{#1}^{\mathrm{(eff)}}}\xspace}        
\def\Cpeff#1  {\ensuremath{\mathcal{C}_{#1}^{'\mathrm{(eff)}}}\xspace}       
\def\Ope#1    {\ensuremath{\mathcal{O}_{#1}}\xspace}                       
\def\Opep#1   {\ensuremath{\mathcal{O}_{#1}^{'}}\xspace}                    

\newcommand{\aunit}[1]{\ensuremath{\text{\,#1}}}       

\newcommand{\tev}{\aunit{Te\kern -0.1em V}\xspace}
\newcommand{\gev}{\aunit{Ge\kern -0.1em V}\xspace}
\newcommand{\mev}{\aunit{Me\kern -0.1em V}\xspace}
\newcommand{\kev}{\aunit{ke\kern -0.1em V}\xspace}
\newcommand{\ev}{\aunit{e\kern -0.1em V}\xspace}
 
\newcommand{\mevc}{\ensuremath{\aunit{Me\kern -0.1em V\!/}c}\xspace}
\newcommand{\gevc}{\ensuremath{\aunit{Ge\kern -0.1em V\!/}c}\xspace}
\newcommand{\mevcc}{\ensuremath{\aunit{Me\kern -0.1em V\!/}c^2}\xspace}
\newcommand{\gevcc}{\ensuremath{\aunit{Ge\kern -0.1em V\!/}c^2}\xspace}

\def\fb   {\ensuremath{\aunit{fb}}\xspace}
\def\invfb   {\ensuremath{\fb^{-1}}\xspace}

\newcommand{\chisq}{\ensuremath{\chi^2}\xspace}

\def\gsim{{~\raise.15em\hbox{$>$}\kern-.85em
          \lower.35em\hbox{$\sim$}~}\xspace}
\def\lsim{{~\raise.15em\hbox{$<$}\kern-.85em
          \lower.35em\hbox{$\sim$}~}\xspace}

\def\sqs   {\ensuremath{\protect\sqrt{s}}\xspace}

\def\pt         {\ensuremath{p_{\mathrm{T}}}\xspace}

\def\tell1  {TELL1\xspace}
\def\ukl1   {UKL1\xspace}

\newcommand{\eg}{\mbox{\itshape e.g.}\xspace}
\newcommand{\ie}{\mbox{\itshape i.e.}\xspace}

\def\be{\begin{equation}}
\def\ee{\end{equation}}
\def\beq{\begin{eqnarray}}
\def\eeq{\end{eqnarray}}

\def\Dst{\Dstarp\xspace}

\newcommand\dmumu{\decay{\Dz}{\mup\mun}}
\newcommand\dpipi{\decay{\Dz}{\pip\pim}}
\newcommand\dkpi{\decay{\Dz}{\Km\pip}}

\newcommand\dhh{\decay{\Dz}{h^{+} h^{-}}}
\newcommand\dstdpi{\decay{\Dstarp}{\Dz \pip}}

\newcommand\dstdkpi{\decay{\Dstarp}{\decay{\Dz(}{\Km\pip)} \pip }}

\newcommand\dgammagamma{\decay{\Dz}{\gamma\gamma}}

\newcommand\dpipipi{\decay{\Dp}{\pip \pim \pip}}
\newcommand\dspipipi{\decay{\Ds}{\pip \pim \pip}}

\newcommand{\bujpsik}{\decay{\Bu}{\jpsi \Kp}}
\newcommand{\bujpsikmumu}{\decay{\Bu}{\decay{\jpsi(}{\mup \mun)}\Kp}}

\def\deltam{\ensuremath{\Delta m}\xspace}
\def\md{\ensuremath{m(\Dz)}\xspace}
\def\mmumu{\ensuremath{m(\mumu)}\xspace}

\newcommand{\runone}{Run~1\xspace}
\newcommand{\runtwo}{Run~2\xspace}

\def\ipchisq{\ensuremath{\rm{IP}\chi^2}\xspace}

\def\brdpipi{\ensuremath{(1.490 \pm 0.027 )\times 10^{-3}}\xspace}
\def\brdkpi{\ensuremath{(3.999 \pm 0.045) \times 10^{-2}}\xspace}

\def\limitoldninety{\ensuremath{6.2 \times 10^{-9}}\xspace} 
\def\limitoldninetyfive{\ensuremath{7.6 \times 10^{-9}}\xspace} 

\def\brlimitninety{\ensuremath{3.1 \times 10^{-9}}\xspace} 
\def\brlimit{\ensuremath{3.1\,(3.5) \times 10^{-9}}\xspace} 
\def\brexplimit{\ensuremath{1.9\,(2.3) \times 10^{-9}}\xspace} 

\def\alphafull{\ensuremath{(2.15 \pm 0.34)\times 10^{-11}}\xspace}

\def\brcentral{\ensuremath{(1.7 \pm 1.0)\times 10^{-9}}\xspace}
\def\nsignalobs{\ensuremath{79 \pm 45}\xspace}
\def\pvalueobs{\ensuremath{0.068}\xspace}
\def\signobs{\ensuremath{1.5\sigma}\xspace}

\usepackage{cite} 
\usepackage{mciteplus}

\usepackage{longtable} 

\begin{document}

\renewcommand{\thefootnote}{\fnsymbol{footnote}}
\setcounter{footnote}{1}


\begin{titlepage}
\pagenumbering{roman}

\vspace*{-1.5cm}
\centerline{\large EUROPEAN ORGANIZATION FOR NUCLEAR RESEARCH (CERN)}
\vspace*{1.5cm}
\noindent
\begin{tabular*}{\linewidth}{lc@{\extracolsep{\fill}}r@{\extracolsep{0pt}}}
\ifthenelse{\boolean{pdflatex}}
{\vspace*{-1.5cm}\mbox{\!\!\!\includegraphics[width=.14\textwidth]{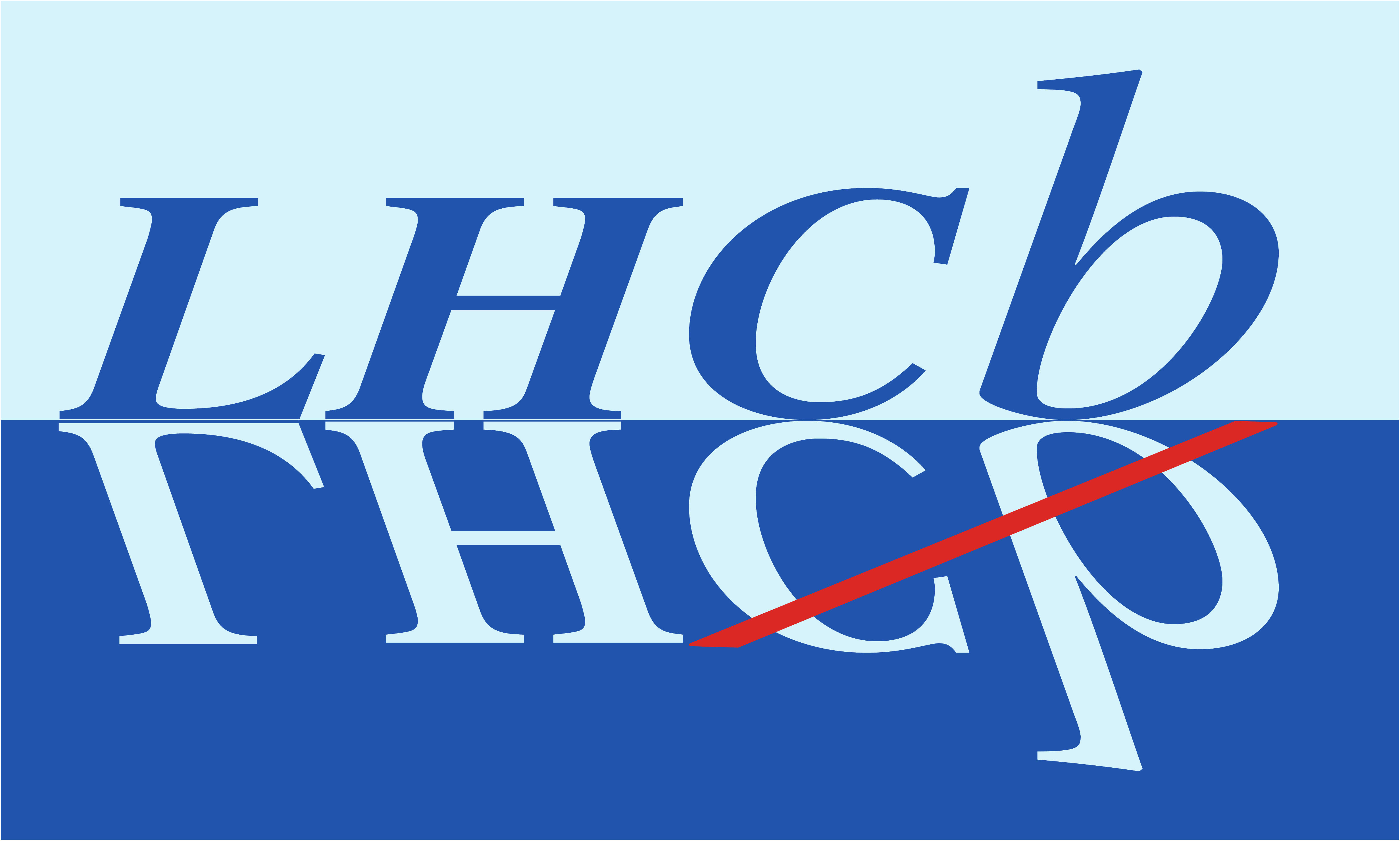}} & &}%
{\vspace*{-1.2cm}\mbox{\!\!\!\includegraphics[width=.12\textwidth]{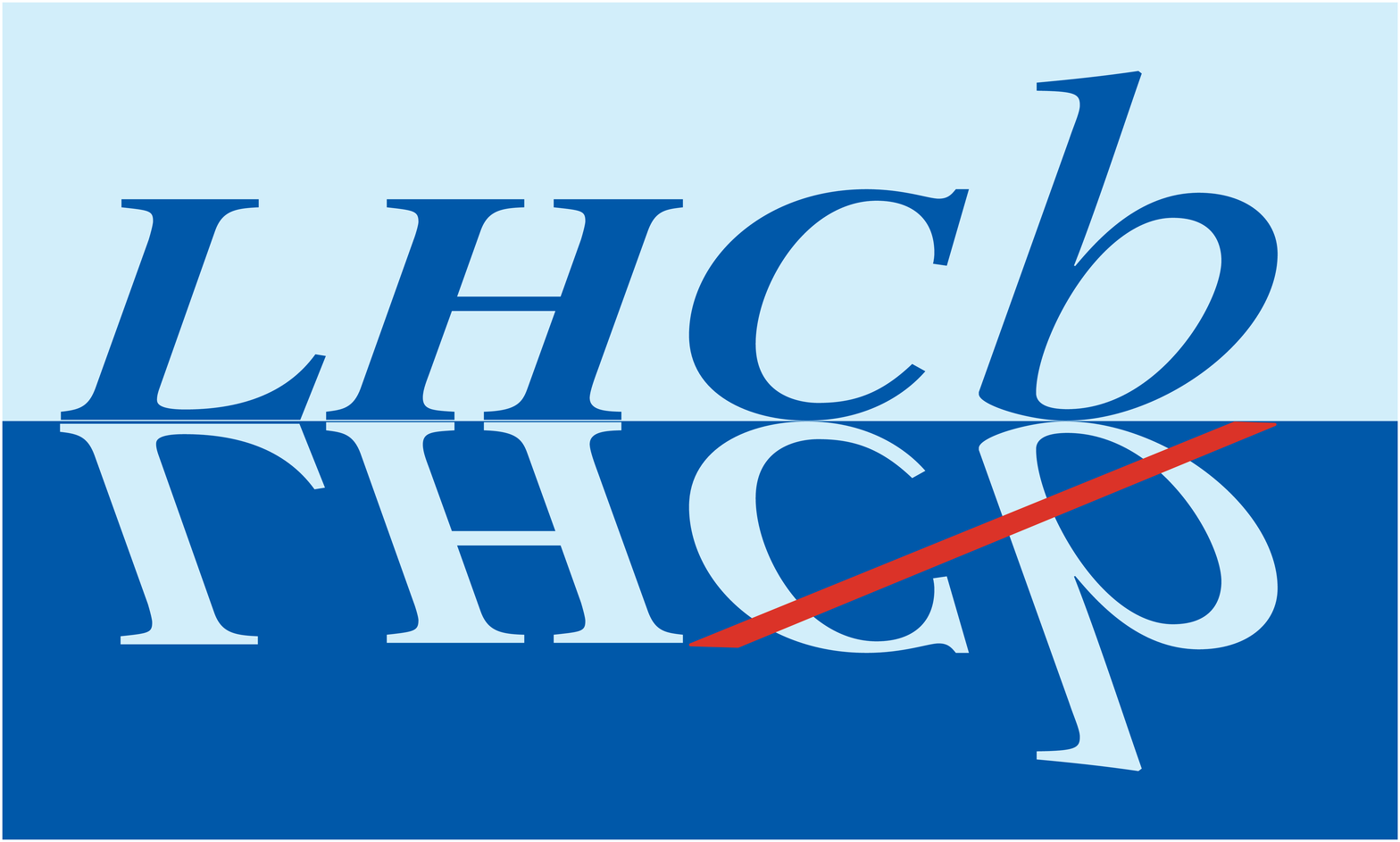}} & &}%
\\
 & & CERN-EP-2022-273 \\  
 & & LHCb-PAPER-2022-029 \\  
 & & November 14, 2023  \\ 
 & & \\
\end{tabular*}

\vspace*{4.0cm}

{\normalfont\bfseries\boldmath\huge
\begin{center}
  \papertitle 
\end{center}
}

\vspace*{2.0cm}

\begin{center}
\paperauthors
\end{center}

\vspace{\fill}

\begin{abstract}
  \noindent
  A search for the very rare \dmumu decay is performed using 
  data collected by the LHCb experiment in proton-proton collisions 
  at $\sqs = 7$, 8 and $13~\tev$, corresponding to an integrated luminosity of $9\invfb$. 
  The search is optimised for \Dz mesons from  \dstdpi decays but is also sensitive to \Dz mesons from other sources.
  No evidence for an excess of events over the expected background is observed. 
  An upper limit on the branching fraction of this decay is set at $\mathcal{B}(\dmumu) < \brlimitninety$ at a 90\% CL. 
  This represents the world's most stringent limit, constraining models of physics beyond the Standard Model. 

\end{abstract}

\vspace*{2.0cm}

\begin{center}
Published in Phys.~Rev.~Lett. 131 (2023) 041804
\end{center}

\vspace{\fill}

{\footnotesize 
\centerline{\copyright~\papercopyright. \href{\paperlicenceurl}{\paperlicence}.}}
\vspace*{2mm}

\end{titlepage}


\newpage
\setcounter{page}{2}
\mbox{~}


\renewcommand{\thefootnote}{\arabic{footnote}}
\setcounter{footnote}{0}

\cleardoublepage


\pagestyle{plain} 
\setcounter{page}{1}
\pagenumbering{arabic}


Processes with a change in quark flavour without a change in electric charge are forbidden at the lowest order in the Standard Model (SM) of particle physics.
Flavour Changing Neutral Currents (FCNC) are  additionally suppressed by the Glashow-Iliopoulos-Maiani (GIM) mechanism~\cite{Glashow:1970gm}. 
FCNC have been extensively studied in strange- and beauty-quark hadrons. 
In the charm sector the GIM suppression is stronger  because the mass differences between down-type quarks are smaller than the ones between up-type quarks. 
These processes can be enhanced by several orders of magnitude in new physics (NP) scenarios when compared to the SM.  

The \dmumu decay is among the most interesting charm-hadrons decays,\footnote{Charge conjugate processes are implied throughout.} being fully leptonic and additionally suppressed by helicity reasons. 
Its SM short-distance contribution is extremely suppressed, yielding a branching fraction on the order of $10^{-18}$~\cite{Burdman:2001tf}. 
Long-distance contributions dominate through an intermediate two-photon state, and can be estimated to be 
 ${\cal B} (\dmumu)\simeq  2.7 \times 10^{-5} {\cal B} (\dgammagamma)$, leading to a \dmumu branching fraction of at least $3 \times 10^{-13}$~~\cite{Burdman:2001tf}.
The best upper limit on the \dgammagamma decay rate was set by the Belle collaboration to be $8.5 \times 10^{-7}$ at a 90\% C.L.~\cite{Nisar:2015gvd}; using the same relation this turns into an upper limit on the long-distance contribution to the \dmumu branching fraction of $2.3 \times 10^{-11}$. 
The \dmumu decay rate can be enhanced in many NP models~\cite{Golowich:2009ii}.
Being one of the most sensitive FCNC processes in the up-quark sector, its branching fraction is used as a primary building block of different models, 
constraining the relevant couplings saturating the branching fraction limit~\cite{Bauer:2015knc,PhysRevD.79.017502,Becirevic:2016oho,Cai:2017wry,Kowalska:2018ulj,Bigaran:2019bqv,Dorsner:2019itg,Kowalska:2020gie,Bordone:2020lnb,Altmannshofer:2020axr,Ishiwata:2015cga,Bharucha:2020eup}.
Most significantly, model-independent bounds on the Wilson coefficients related to charm physics and in particular to \dmumu decays have been set in Ref.~\cite{Fajfer:2015mia,Bharucha:2020eup}.
Furthermore, the \dmumu rate is correlated to the rate of $\Dz - \Dzb$ mixing in many NP models~\cite{Golowich:2009ii}. 
This is of uttermost importance given the recent first observation 
of the mass difference between neutral charm-meson eigenstates~\cite{LHCb-PAPER-2021-009}.
Concerning specific models, in the Minimal Supersymmetric Standard Model no sizeable contribution would enhance the \dmumu rate~\cite{Golowich:2009ii}. Conversely, in some Supersymmetric models with $R$-parity symmetry violation tree level contributions would be allowed~\cite{Burdman:2001tf,Burdman:2003rs}. 
Recent discussions of predictions for \dmumu decays in such models can be found in Refs.~\cite{Wang:2014uiz,Altmannshofer:2020axr}.
In addition, it is interesting to note the importance of \dmumu decays as testing ground 
for models with leptoquarks proposed to explain deviations from the SM observed in \B physics measurements~\cite{LHCb-PAPER-2017-013,LHCb-PAPER-2021-004, LHCb-PAPER-2019-040, LHCb-PAPER-2020-041, LHCb-PAPER-2020-002, LHCb-PAPER-2015-051, LHCb-PAPER-2021-022, ATLAS:2018gqc, BaBar:2006tnv, BaBar:2015wkg, Belle:2009zue, Belle:2016fev, CDF:2011tds, CMS:2015bcy, CMS:2017rzx, LHCb-PAPER-2016-012, LHCb-PAPER-2021-014, LHCb-PAPER-2014-006, LHCb-PAPER-2015-009,LHCb-PAPER-2022-045}.
In some of these models rare \B decays receive contributions at the loop level, while new particles could mediate the \dmumu decay at tree level. This phenomenon has been extensively discussed in the 
literature~\cite{Bauer:2015knc,Freytsis:2015qca,Benbrik:2008ik,PhysRevD.79.017502,deBoer:2015boa,Becirevic:2016oho,Cai:2017wry,Kowalska:2018ulj,Bigaran:2019bqv,Dorsner:2016wpm,Dorsner:2019vgp,Dorsner:2019itg,Kowalska:2020gie,Bordone:2020lnb,Abudinen:2022dnw}. 
In those where additional vector bosons ($Z^\prime$) are introduced,  the \dmumu decay usually does not give strong constraints~\cite{Kang:2019vng,Alok:2019xub}. 
Similarly, in other explanatory models ~\cite{Guadagnoli:2018ojc,Alonso:2018bcg,Alok:2015iha} the bound from \dmumu decays can be avoided.
Instead in some models with vector-like fermions~\cite{Ishiwata:2015cga}, the \dmumu decay gives the strongest constraint.

The current world's best limit on this decay is $\mathcal{B}(\dmumu ) < \limitoldninety (\limitoldninetyfive)$ at 90\% (95\%) CL, and was obtained by the LHCb experiment exploiting about 0.9~\invfb of 2011 data~\cite{LHCb-PAPER-2013-013}. 
The data used in Ref.~\cite{LHCb-PAPER-2013-013} are also used in this analysis and those results are superseded by this Letter.

This Letter presents a search for the \dmumu decay based on data collected by the LHCb experiment in \proton\proton collisions corresponding to $9\invfb$ of integrated luminosity. 
The data have been  collected in 2011, 2012 (\runone) and 2015--2018 (\runtwo) at $\sqrt{s} = 7$, 8 and $13\tev$, respectively. Compared to the previous publication~\cite{LHCb-PAPER-2013-013}, the present work benefits from  various improvements in the analysis, such as refined multivariate algorithms against combinatorial and misidentified background as well as an improved trigger~\cite{LHCb-DP-2019-001}, 
described throughout the paper. These allow to mitigate the impact of the harsher experimental conditions in \runtwo. Both the higher energy and instantaneous luminosity produce a higher track multiplicity, which increases the combinatorial background and worsens the particle identification performance.
The \dmumu decay is searched for using \dstdpi decays, as this improves the background rejection and allows the yield of the decay to be obtained from a two-dimensional fit to the dimuon invariant mass, \mmumu,  and the difference between the $\Dst$ and $\Dz$ candidate masses, \deltam. The yield is converted to the decay branching fraction by normalising to two hadronic decays, \dkpi and \dpipi, selected concurrently to the signal (collectively referred to as \dhh). 

The \lhcb detector is a single-arm forward spectrometer covering the pseudorapidity range \mbox{$2<\eta<5$}, described in detail in Refs.~\cite{LHCb-DP-2008-001,LHCb-DP-2014-002}. 
The simulated events used in this analysis are produced with the software described in Refs.~\cite{Sjostrand:2007gs,*Sjostrand:2006za,LHCb-PROC-2010-056,Lange:2001uf,Agostinelli:2002hh,Allison:2006ve,LHCb-PROC-2011-006,davidson2015photos}. 

Events are selected online by a trigger that consists of a hardware stage, which is based on information from the calorimeter and muon systems, followed by two software stages.
At the hardware trigger stage, events are required to have a muon candidate with high transverse momentum, \pt, or a hadron, photon or electron candidate with high transverse energy in the calorimeters.
A first stage of the software trigger selects events with a muon candidate, or a high \pt charged particle, or a combination of two tracks, each of these displaced from the primary $pp$ collision vertex (PV). 
In the second stage of the software trigger, dedicated algorithms select candidate \dmumu, \dkpi and \dpipi decays, combining two oppositely charged tracks with loose particle identification (PID) requirements that form a secondary vertex separated from any PV. The invariant mass of the \Dz candidate must lie in an interval of $\pm 300 ~(\pm 70)~\mevcc$ centred on the  known \Dz mass for signal (normalisation) channel candidates. 
To keep the same trigger selection for the signal and normalisation channels, aside from PID, a scale factor of order 0.2--3.0\% depending on the data taking period is applied to the normalisation channels trigger selections to limit their rate, keeping randomly a fraction of the events.
 In the offline selection, only candidates associated with a muon hardware trigger, or those where the rest of the event contained a high transverse energy hadron or electron are kept for the signal channel. For the normalisation channels, only candidates associated with a high transverse energy hadron are kept.
 
 In the offline analysis, \Dz candidates satisfying the trigger requirements are formed with similar but more stringent criteria than the second  software-trigger stage. These \Dz candidates are then combined with a charged particle originating from the same PV and having $\pt > 110~\mevc$ to form $\Dstar^+ \to \Dz \pi^+$ candidates. 
 To improve the mass resolution, the $\Dst$ meson decay vertex is constrained to coincide with the PV~\cite{Hulsbergen:2005pu}. The candidate \deltam is required to be in the range 139.6–151.6~\mevcc. A multivariate selection based on a boosted decision tree (BDT) algorithm~\cite{Breiman,Adaboost,Hocker:2007ht} is used to suppress background from random combinations of charged particles, using as input: the \pt of the pion from the \Dst decay, the smallest \pt and impact parameter significance with respect to the PV of the \Dz decay products, the angle between the \Dz momentum and the vector connecting the primary and secondary vertices, and the quality of the \Dz vertex. The BDT is trained separately for each Run of data taking, using simulated decays as signal and data candidates from the dimuon sample with $\mmumu > 1894~\mevcc$ as background. 
 The $k$-folding technique, with $k=9$, is applied~\cite{Blum:1999:BHB:307400.307439}. 
 The BDT output ranges from 0 to 1, from background-like candidates to more  signal-like candidates.
 The BDT output is used to define three search regions: $\text{BDT} \in [ 0.15, 0.33]$, $[0.33, 0.66]$, $[0.66,1.]$. 
 The output of the same BDT algorithm is computed also for the \dpipi candidates for calibration purposes, but is not required in the selection.  

A second source of background is due to two- and three-body \Dz decays, with one or two hadrons misidentified as muons (\eg \dhh or semileptonic decays). The misidentification occurs mainly for hadrons that decay into a muon before the muon sub-detector. Although this process is relatively rare, the large branching fractions of these modes produce a background peaking in the signal region of the \mmumu distribution that is partially suppressed by a multivariate muon identification discriminant combining information from the Cherenkov detectors, the calorimeters, and the muon sub-detector~\cite{LHCb-DP-2018-001}. In addition, the muon candidates are required to have associated muon chamber hits that are not shared with any other track in the event. 
Signal and background sources can originate both from  \dstdpi decays or from other sources (the PV or \B decays) combined with an unrelated pion (untagged); both are taken into account in the yield estimation.
A requirement on the output of the multivariate muon identification discriminant is simultaneously chosen for the three BDT regions, by optimising the sensitivity to the minimum visible cross-section, as defined by an extension of the figure of merit defined in Ref.~\cite{Punzi:2003bu}.
Roughly one percent of events contain more than one signal candidate after all selection requirements, all of which are retained.

The signal yield is converted to the decay branching fraction by normalising to the hadronic decays \dpipi and \dkpi, with branching fractions of \brdpipi and \brdkpi, respectively~\cite{HFLAV18}, as 
\begin{equation}
\mathcal{B}(\dmumu) = \frac{{N}_{\dmumu}}{N_{\dhh}} \cdot
\frac{\varepsilon_{h^+h^-} }{\varepsilon_{\mu^+\mu^-}}\cdot s \cdot \mathcal{B}(\dhh)  \equiv \alpha N_{\dmumu}
\label{eq:br_norm}
\end{equation}
where $\varepsilon$ is the efficiency and $N$ is the yield of the given channel, $s$ is the scale factor of the normalisation channel and $\alpha$ is defined as the single event sensitivity.

The efficiencies in Eq.~\eqref{eq:br_norm} are factorised into different steps for ease of estimation and evaluated with respect to the previous steps: detector acceptance, reconstruction and selection, PID, and trigger.

The reconstruction and selection efficiencies are obtained from simulated samples.
 The simulated candidates are assigned weights with an iterative procedure that improves the agreement with data using the following variables: pseudorapidity of the \Dz meson, transverse momentum of the \Dz meson and number of tracks in the event. It is verified that after weighting, all variables used in the selection agree well between data and simulation.
The weights obtained from the \dpipi candidates are used to correct also the signal simulation.

 Possible residual differences between data and simulation in the tracking efficiencies are determined using control channels in data~\cite{LHCb-DP-2013-002}. 
The PID efficiencies are determined from data using samples of kinematically identified charged particles from \bujpsik and \dstdkpi decays~\cite{LHCb-DP-2018-001}, weighted to match the kinematic properties of the signal and the normalisation channels, respectively. The efficiencies are determined in bins of 
the $p$  and \pt of the tracks. A total systematic uncertainty of 1--3\% is associated to the binning scheme and background determination in the calibration samples. 

The efficiency of the second level of the software trigger is unity with respect to the offline-selected candidates by construction, as the selection is tighter in every requirement.
The hardware and first level software trigger efficiencies are evaluated with the \textsc{Tistos} method~\cite{Tolk:1701134} in data.  
For the signal channel, the \bujpsikmumu decay is used as the calibration channel, selected with the same requirements as those used for the analysis of \B decays into two muons~\cite{LHCb-PAPER-2021-007,LHCb-PAPER-2021-008}.
The calibration is performed in intervals of the \jpsi \pt and pseudorapidity.
For each interval, a scaling factor between data and simulation is obtained and applied to the \dmumu simulation.  Compatible results are obtained repeating the calibration in intervals of \pt and the maximum \ipchisq of the muons, where \ipchisq is defined as 
the difference in the vertex-fit \chisq of a given PV reconstructed with and
without the track under consideration. The typical scaling between data and simulation deviates from unity by 2--6\%. 
The normalisation channels, given their high yields, are self-calibrated.
The \textsc{Tistos} method is applied to the \dkpi and \dpipi channels and trigger efficiencies are obtained.
To minimise cross-correlation biases, only candidates in events that satisfy a muon trigger independently of the candidate are used as calibration sample. 
The calibration of the hadronic hardware trigger is also validated with independent estimates based on control samples in data, obtained with similar methods as in Ref.~\cite{LHCb-PUB-2011-026}, from which a 15\% relative systematic uncertainty is assigned to the hadronic trigger efficiency calibration. 

The efficiency of the BDT requirement, and the signal fraction in the BDT intervals, are calibrated in data by applying the same estimator to the \dpipi decay, which is topologically very similar to the signal. The distribution of the BDT output is obtained in background subtracted \dpipi decays in data and simulation, and found to be compatible, as shown in the Supplemental Material~\cite{supplemental}.
A small correction is determined and applied to the signal; its uncertainty is assigned as systematic uncertainty to the signal efficiency.

\begin{figure}
\includegraphics[width = 0.5\textwidth]{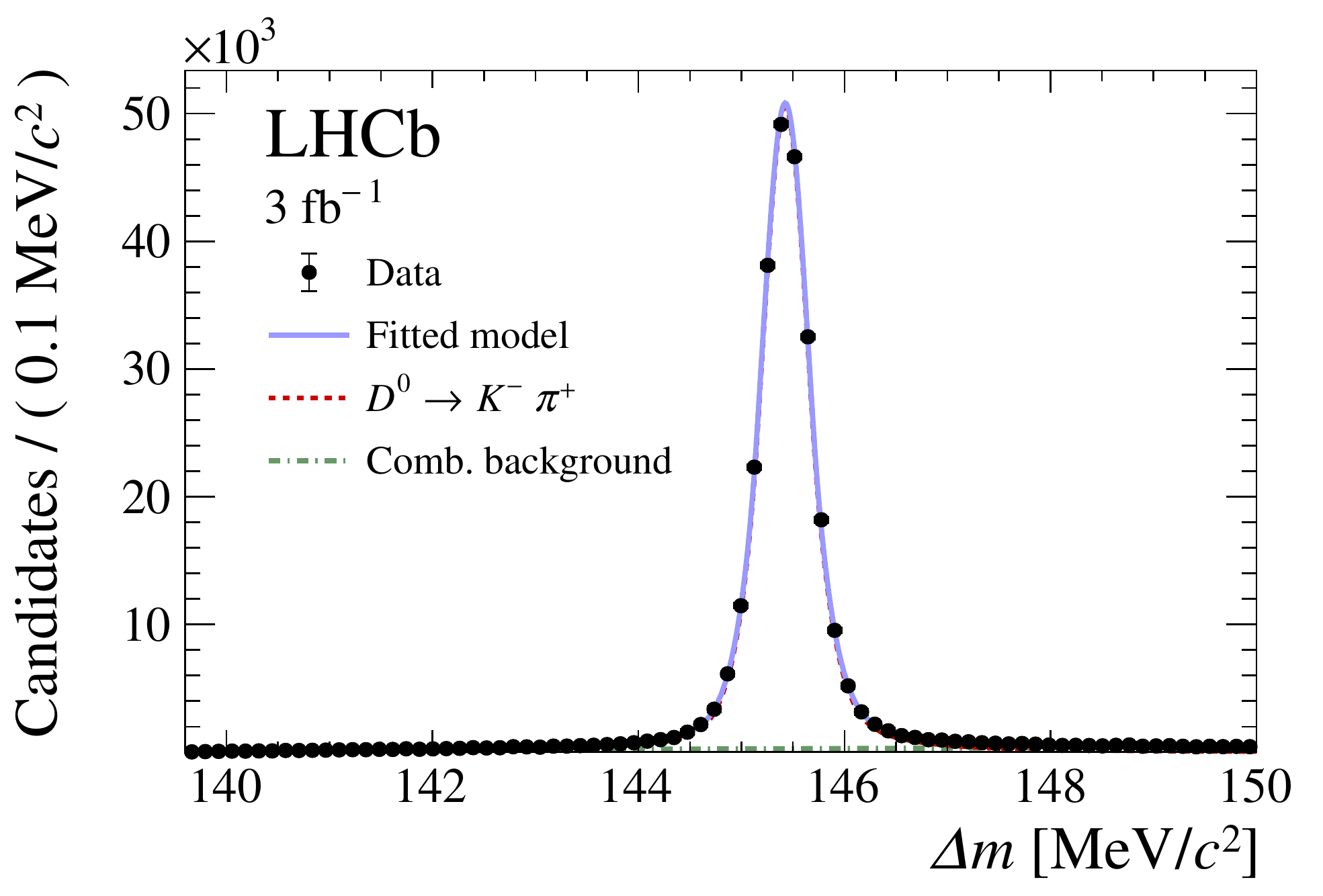}
\includegraphics[width = 0.5\textwidth]{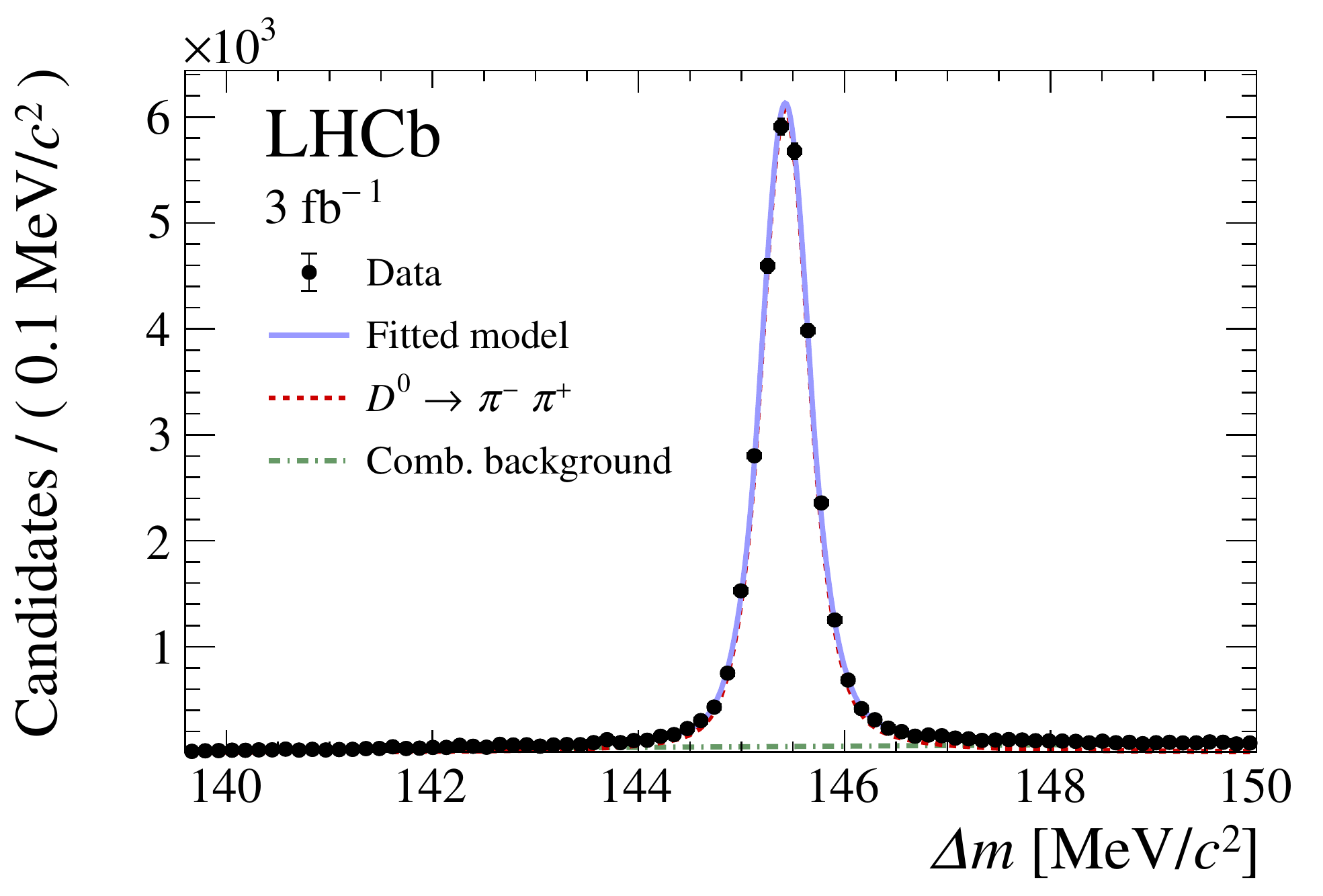}
\includegraphics[width = 0.5\textwidth]{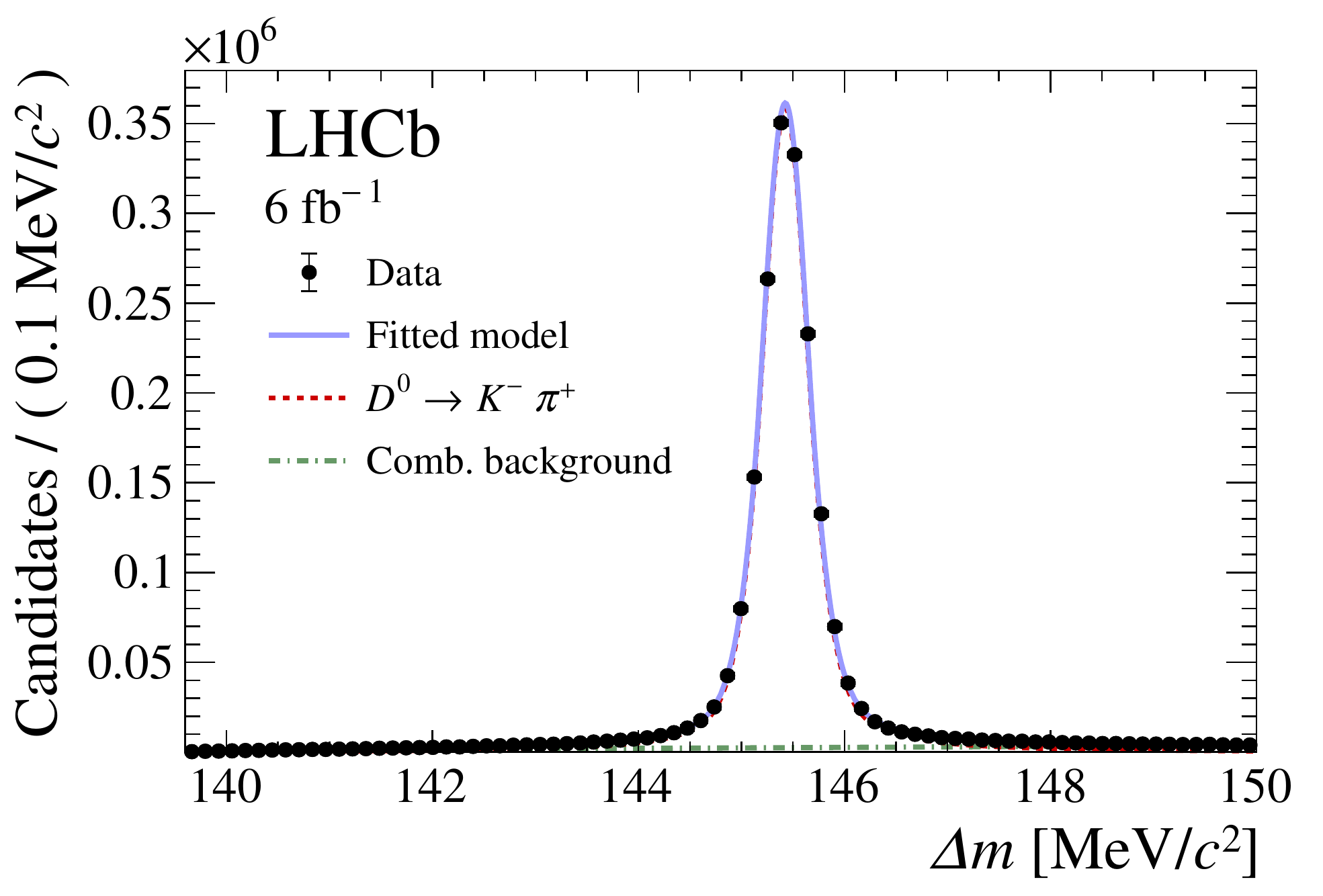}
\includegraphics[width = 0.5\textwidth]{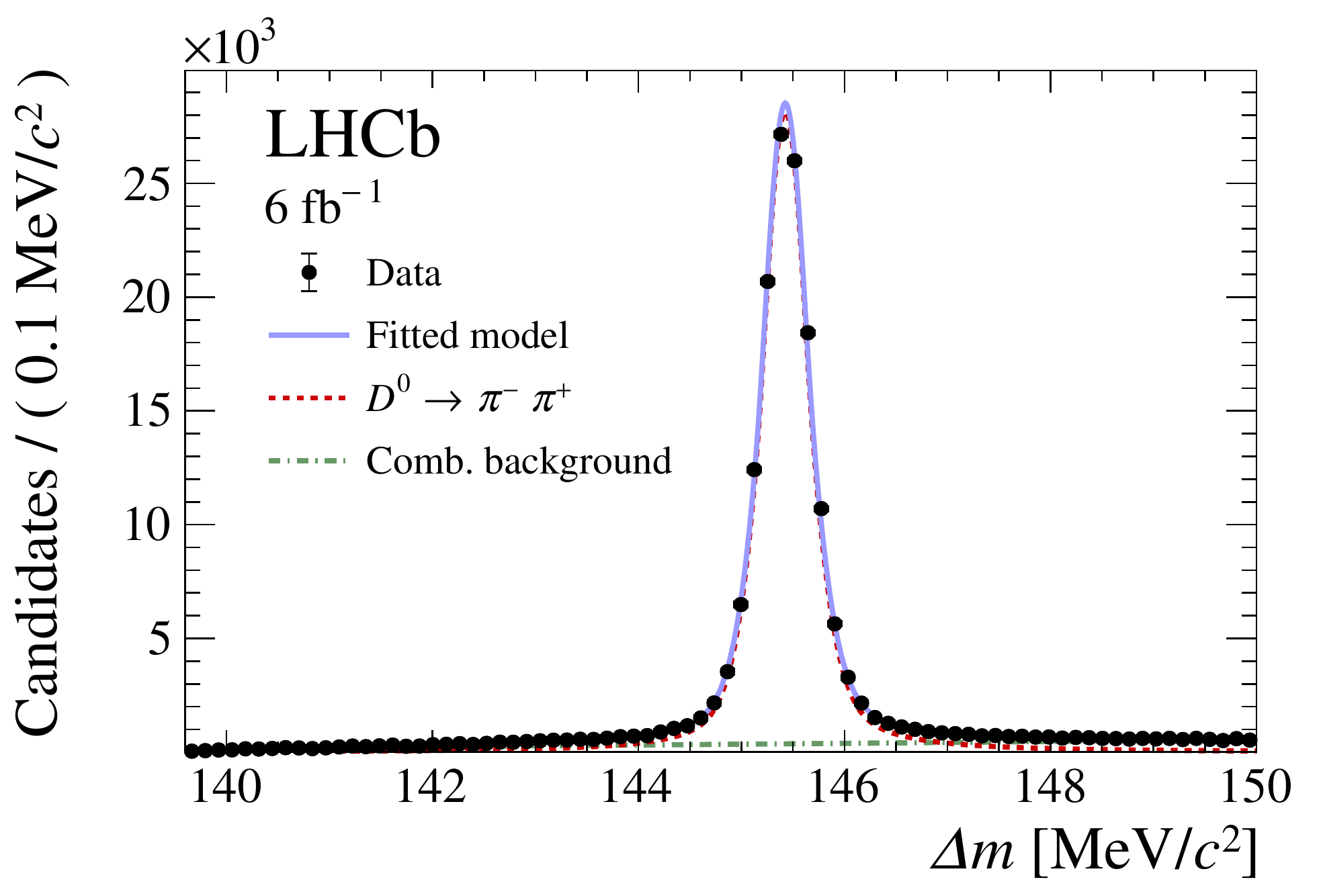}
 \caption{Distributions of \deltam for (left) \dkpi and (right) \dpipi normalisation channels candidates for (top) \runone and (bottom) \runtwo data. 
 The distributions are superimposed with the fit. }\label{fig:normplot}
\end{figure}

The yields of the normalisation channels are obtained through a fit to the \deltam distribution (Fig.~\ref{fig:normplot}), requiring the reconstructed \Dz  mass to be within $\pm 10 \mevcc$ of the known \Dz mass. 
The signal probability distribution function is composed of a sum of a Gaussian 
and a Crystal-Ball function~\cite{Skwarnicki:1986xj} with power-law tails on both sides. 
The background is described with a threshold function, as defined in Ref.~\cite{LHCb-PAPER-2013-013}.
The parameters of the Crystal-Ball function are estimated with simulation: the power of the tail is fixed, while the position where the power tails start may vary freely in the fit. 
In addition, the signal width and all background parameters are left free in the fit.

Using Eq.~\eqref{eq:br_norm}, values of $\alpha$ for both normalisation channels are obtained, and found to be in good agreement with each other for each data taking run and for the full sample. 
As an additional cross-check, the ratio of the efficiency corrected yields of the two normalisation channels is obtained and compared to the ratio of their branching fractions. The value is stable across the data taking years and compatible with the world average~\cite{HFLAV18}. 
The average single event sensitivity is found to be $\alpha = \alphafull$, 
corresponding to at most one expected signal \dmumu decay under the SM hypothesis.

\begin{figure}[tbp]
\includegraphics[width = 0.5\textwidth]{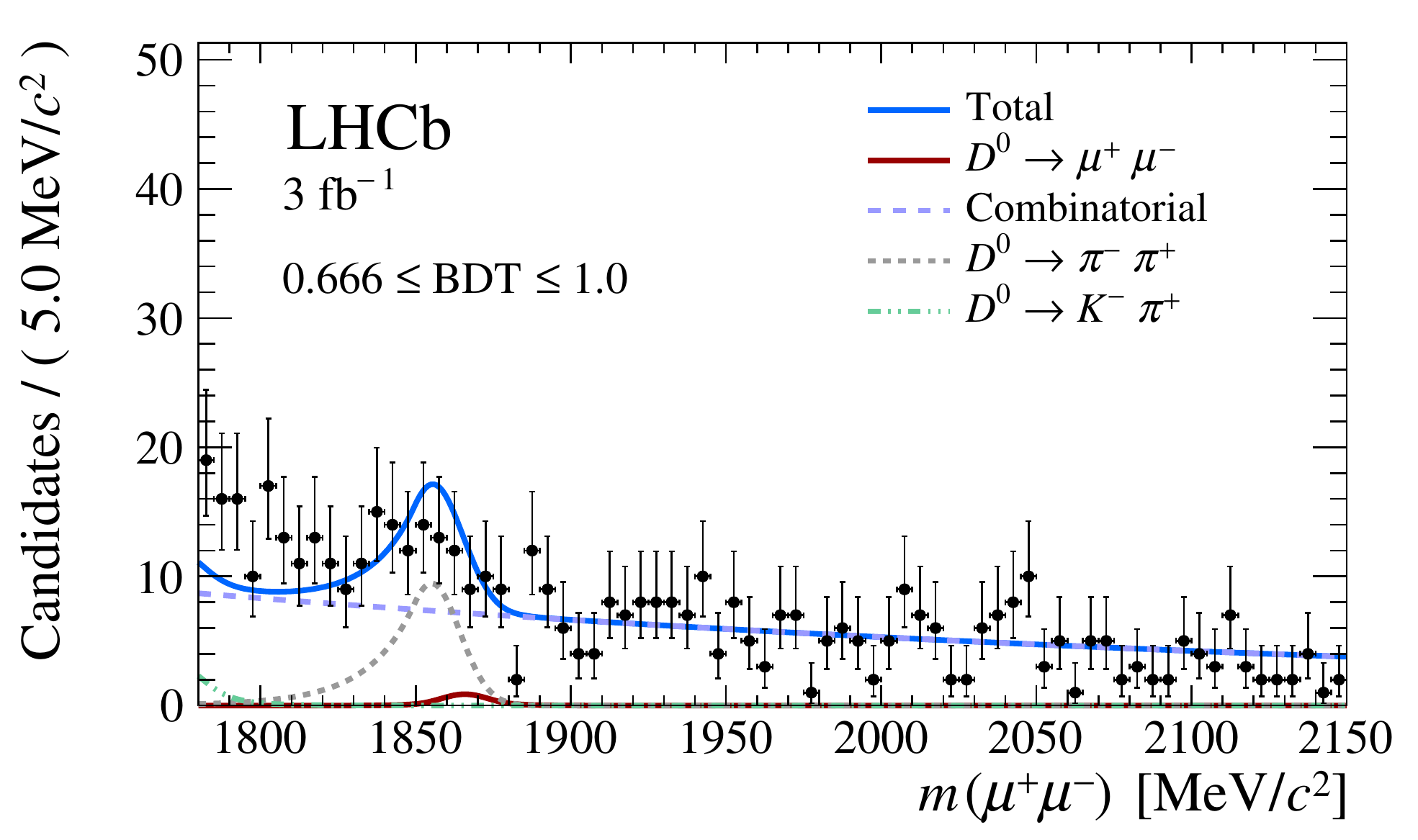}
\includegraphics[width = 0.5\textwidth]{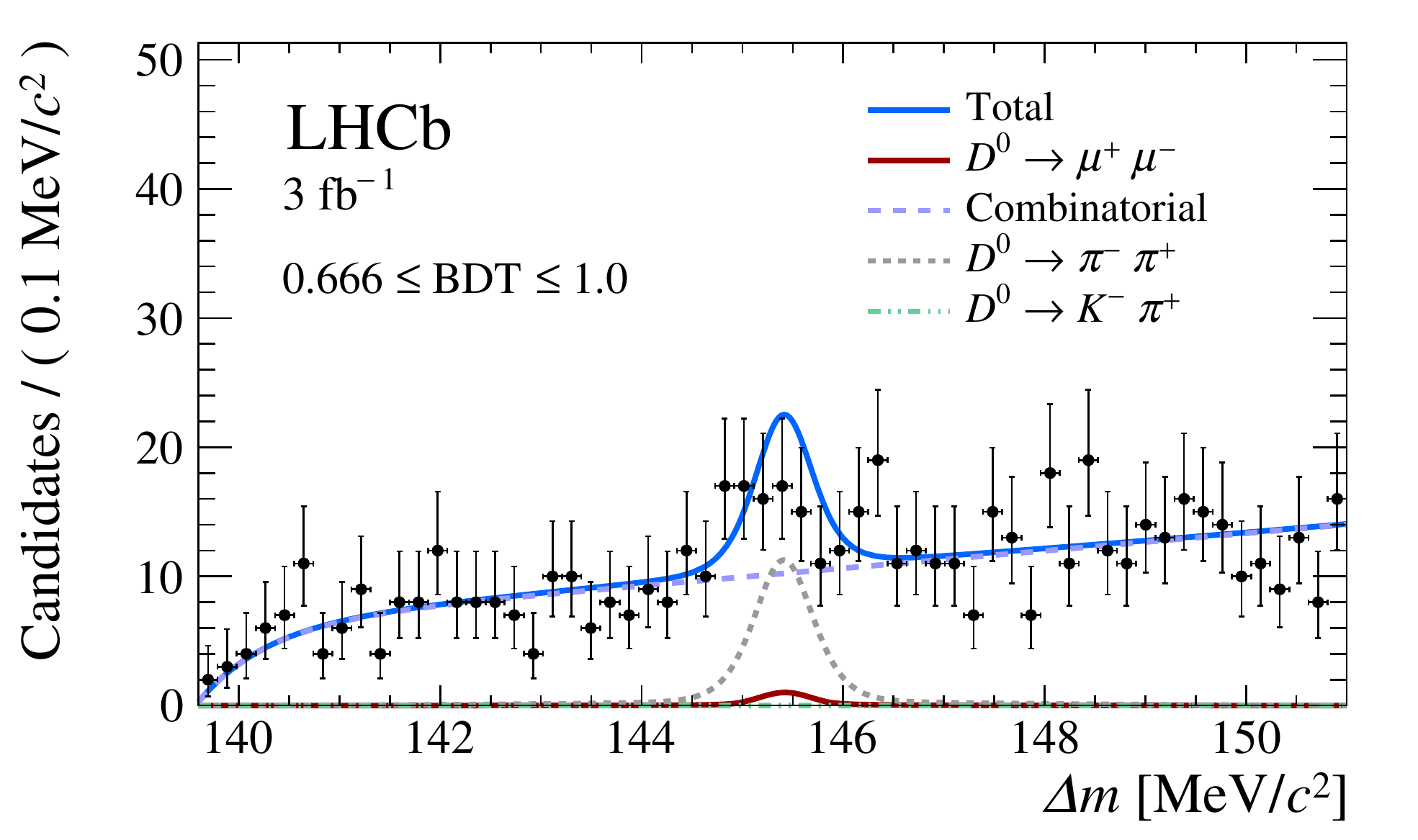}
\includegraphics[width = 0.5\textwidth]{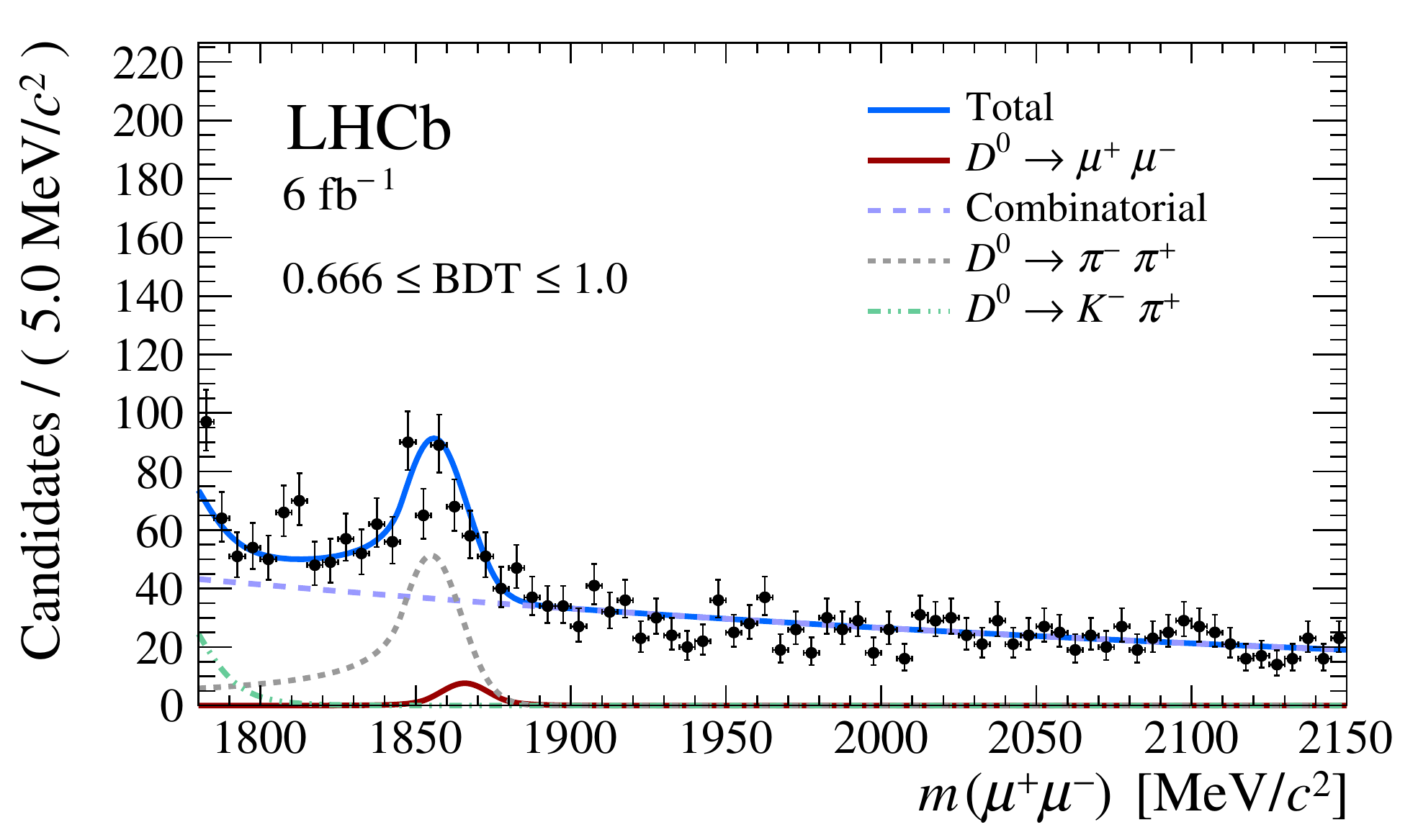}
\includegraphics[width = 0.5\textwidth]{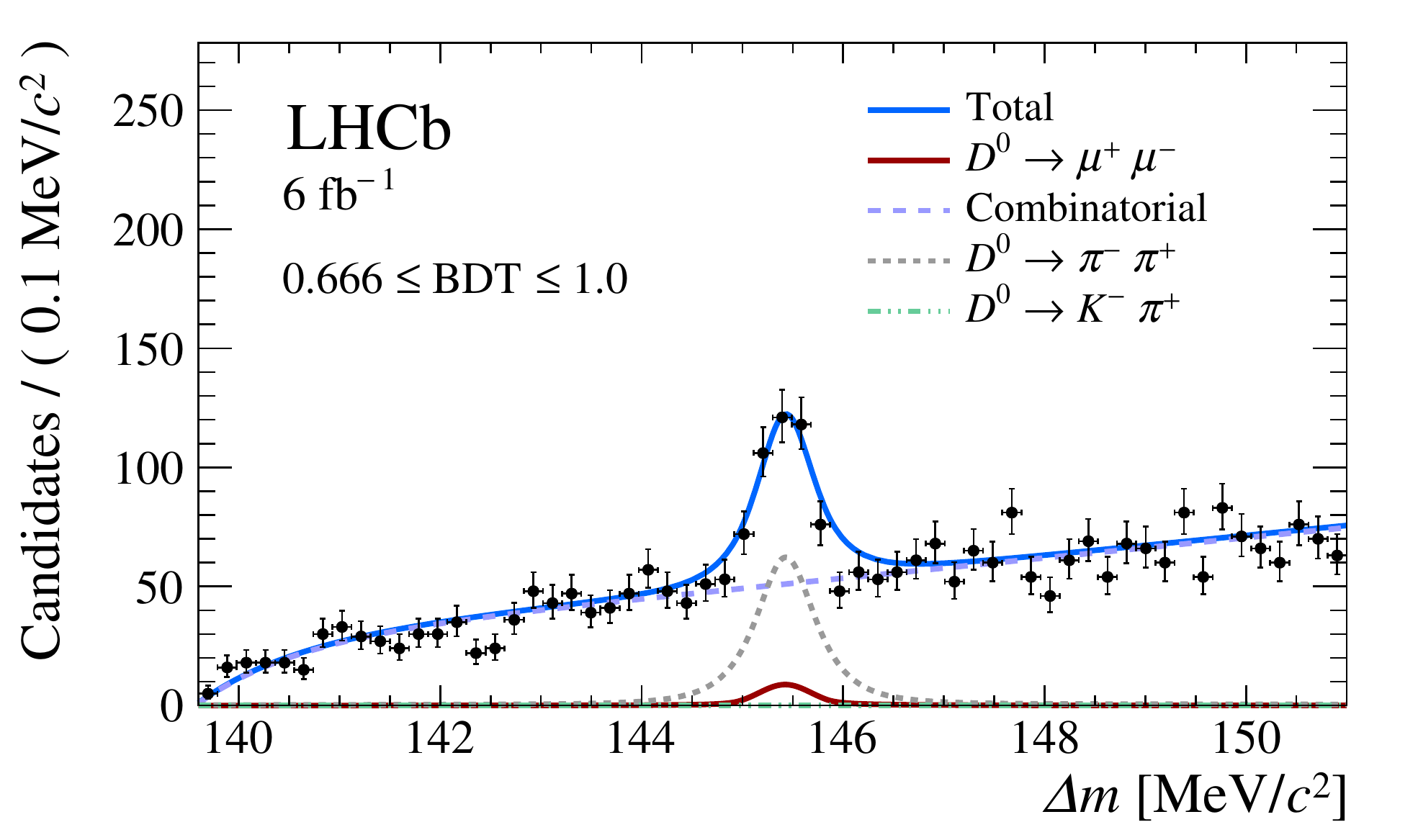}
\caption{Distribution of (left) \mmumu and (right) \deltam  for the \dmumu  candidates in data from (top) \runone and (bottom) \runtwo, for the most sensitive BDT interval.
 The distribution is superimposed with the fit to data. 
  Each of the two distributions is in the signal region of the other variable, see text for details.    
  Untagged and tagged decays are included in a single component for signal and \dpipi background.
 }\label{fig:signalplot}
\end{figure}

The signal yield is obtained through an unbinned maximum-likelihood fit to the two-dimensional distribution of \mmumu and \deltam, performed simultaneously in the three BDT intervals and in the two data taking Runs. 
The distributions projected onto the two variables are shown in Fig.~\ref{fig:signalplot}. Each of the two projections is selected using only candidates in the signal region of the other variable, where the signal regions are defined as 
$m(\Dz) \in [1840,1885]\mevcc$  and $\deltam \in [144.9,146.1]\mevcc$, respectively. The full distributions can be seen in the Supplemental Material~\cite{supplemental}.
The correlation between the two variables is found to be negligible for all contributing decay modes, thus they are treated as uncorrelated.
After the full selection, only combinatorial and misidentified hadronic \Dz decays are found to contribute to the background. Background from semileptonic decays is found to be
negligible, and any remaining background from other sources is well modelled as part of the combinatorial background component.

The shape of the signal and misidentified background (\dpipi, \dkpi) distributions in the two fit variables is obtained from simulation, reconstructed as \mbox{\dmumu} decays. 
The model parameters are determined separately for \runone and \runtwo; the resulting PDF describes the distribution in each data taking year and BDT interval well.
For signal, the \mmumu  and \deltam distributions are both parametrised by Crystal-Ball~\cite{Skwarnicki:1986xj} functions with power tails on both sides. For \dpipi decays, a single Crystal-ball function  in \mmumu and the sum of a Johnson~\cite{Johnson:1949zj} and Gaussian function in \deltam are employed. For \dkpi decays, a Johnson function is used for the \mmumu distribution, while the \deltam distribution is described by three Gaussian functions. The combinatorial background is described by an exponential function in \mmumu and a threshold function in \deltam~\cite{LHCb-PAPER-2013-013}. 
Untagged signal and \dpipi components are included and parametrised as their respective tagged component in \mmumu and with the same threshold function of the combinatorial background in \deltam. The fraction of this component is fixed to the value determined in each BDT interval from a fit to \dpipi data.  The shape parameters obtained from the simulated samples are fixed in the data fit, while the slope of the exponential of the combinatorial background is left free to vary in each BDT interval.

A constraint on the expected number of misidentified \dpipi decays is determined from a dedicated, high-statistics, simulation sample with the trigger and offline selection applied. The most critical part of the simulated sample is the PID efficiency due to the presence of a large fraction of $\pi \to \mu$ decays that mimic the signal and are not considered in the standard calibration tools. The PID efficiency is obtained from simulation but it is cross-checked using $\dpipipi$ and $\dspipipi$ control samples in data where same-sign pions are weighted to match the kinematics of \dpipi decays. The agreement between the PID efficiency determined with both methods is satisfactory over the full range of the muon identification discriminant variable~\cite{supplemental}. Therefore, no systematic uncertainty is assigned on this estimate. The uncertainty on the expected \dpipi yield is propagated through a Gaussian constraint on the relevant parameter in the final fit. 

The yield of the misidentified \dkpi decays is constrained from an auxiliary fit to the \mmumu  sideband data, recomputed with the correct mass hypothesis. The fit is performed using the \dm distribution within a $\pm10 \mevcc$ region around the \Dz mass in the $\Km\pip$ mass hypothesis. A correction is applied to take into account this mass requirement. The correlation between this estimate and the yield in the final fit is found not to influence the estimate of the signal branching fraction.

The systematic uncertainties related to both the normalisation, through $\alpha$, and the background shapes and yields, are included in the fit as Gaussian constraints on the relevant parameters.
The dominant systematic uncertainty comes from the calibration of the hadronic trigger efficiency, which is shared through auxiliary parameters among the normalisation channels, and also with the misidentified \dpipi yields that depend on the same estimate. 
The fit procedure is tested with pseudoexperiments.
The values of the floating shape parameters are obtained from the data fit.  Unbiased estimates of the branching fraction with correct coverage are obtained. 

The \mmumu and \deltam distributions in data are shown for the most sensitive BDT interval in Fig.~\ref{fig:signalplot} and for all intervals in Ref.~\cite{supplemental}, overlaid with   
the result of the fit.
The data are consistent with the expected background. 
The value obtained for the \dmumu branching fraction is
$\mathcal{B}(\dmumu)  =   \brcentral$, corresponding to \nsignalobs signal decays. 
The significance of this signal is estimated comparing the test statistics in data with the distribution of the test statistics in background-only pseudoexperiments, and is found to have a $p$-value of \pvalueobs, corresponding to a significance of \signobs (see also Ref.~\cite{supplemental}).
An upper limit on the branching fraction is derived using the frequentist   $\rm{CL_s}$ method~\cite{Read:2002hq} as implemented in the \textsc{GammaCombo} framework~\cite{GammaCombo,LHCb-PAPER-2016-032}. This yields
\begin{equation*}
    \mathcal{B}(\dmumu) < \brlimit  \text{ at 90 (95)\% CL }.
\end{equation*}
The observed limit is larger than the one expected from background-only pseudoexperiments, $\mathcal{B}(\dmumu) < \mbox{\brexplimit}  \text{ at a 90 (95)\% CL}$, coherently with the central value for the  signal branching fraction. 

The fit is repeated with different configurations: allowing the resolution of the misidentified \dpipi background to vary, using a double exponential function in place of a single one for the combinatorial background, and reducing the range in the \deltam variable. No significant change was found in the signal branching fraction with any configuration. 

In summary, a search for the \dmumu decay in data corresponding to $9\invfb$ of $pp$ collision data collected by the LHCb experiment is performed. No excess with respect to the background expectation has been found and an upper limit of 
$\mathcal{B}(\dmumu) < \brlimitninety$ at 90\% CL has been set. 
This result represents an improvement of more than a factor two with respect to the previous LHCb result. This measurement constitutes the most stringent limit on the relevant FCNC couplings in the charm sector, allowing to set additional constraints on physics models beyond the SM which predict the branching fractions of \dmumu and describe results from $B$ physics measurements.

\addcontentsline{toc}{section}{References}
\bibliographystyle{LHCb}
\bibliography{main,standard,LHCb-PAPER,LHCb-CONF,LHCb-DP,LHCb-TDR}
 
 \cleardoublepage
\section*{Supplemental Material }
\label{sec:supplemental}

\begin{figure}[!bp]
\includegraphics[width = 0.5\textwidth]{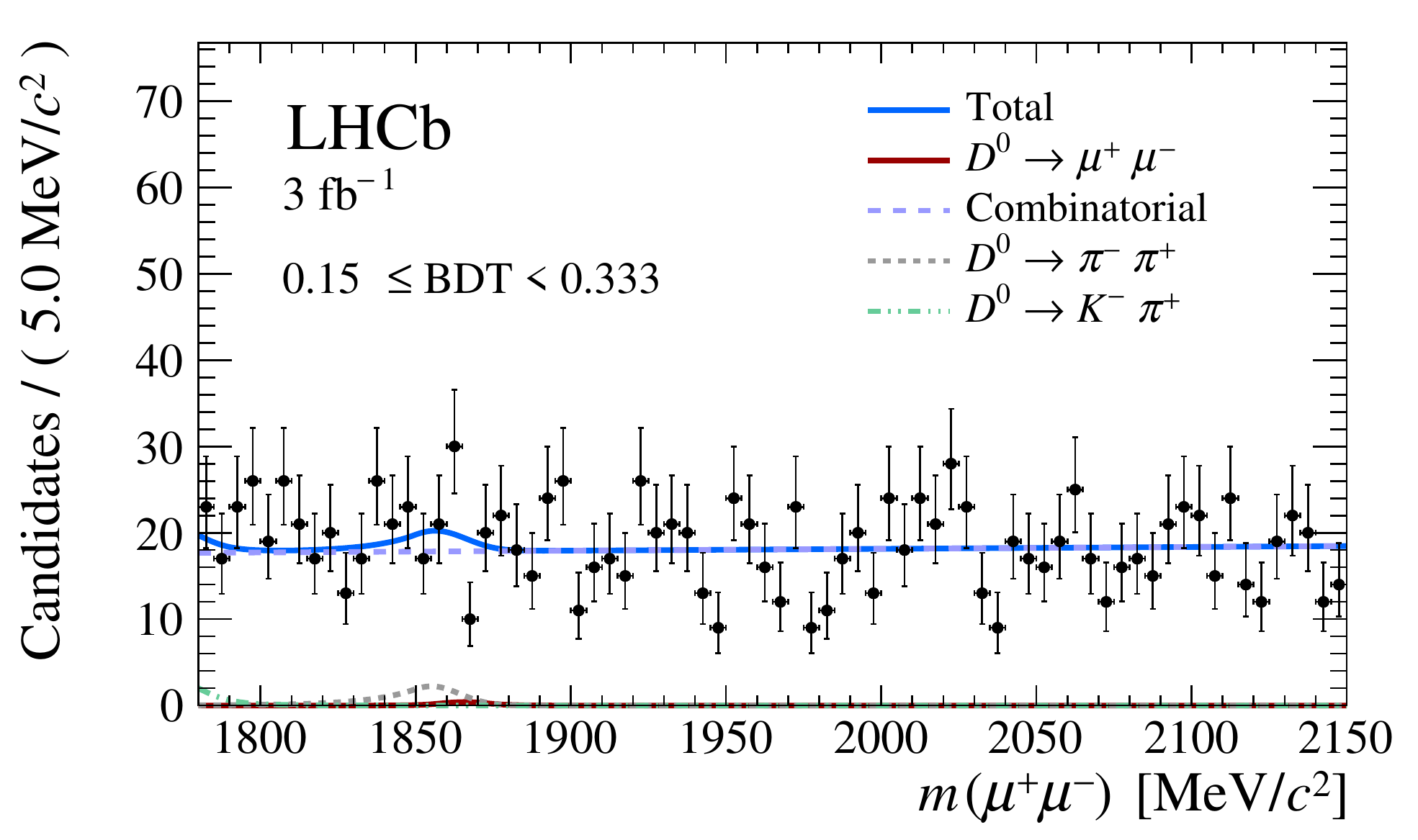}
\includegraphics[width = 0.5\textwidth]{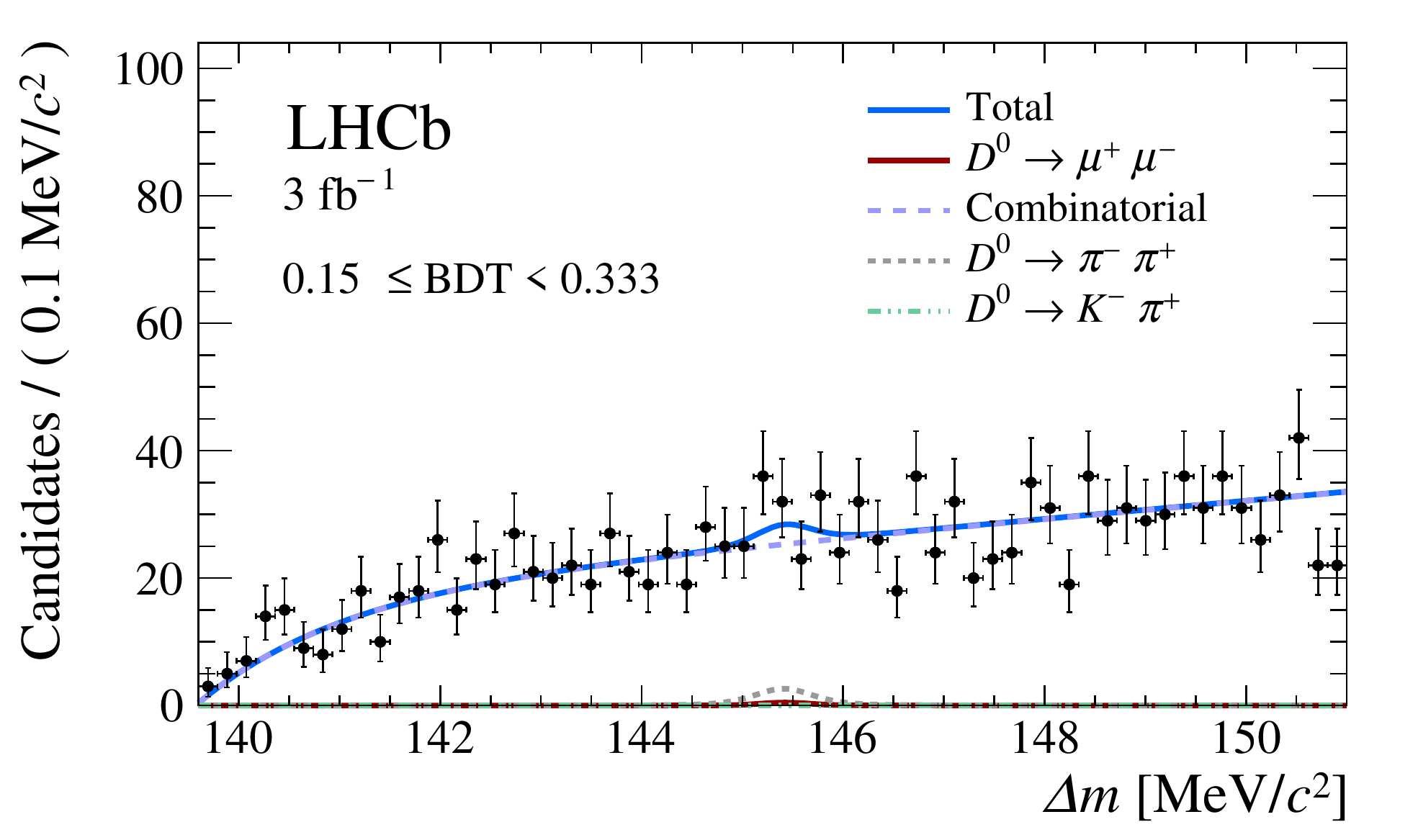}
\includegraphics[width = 0.5\textwidth]{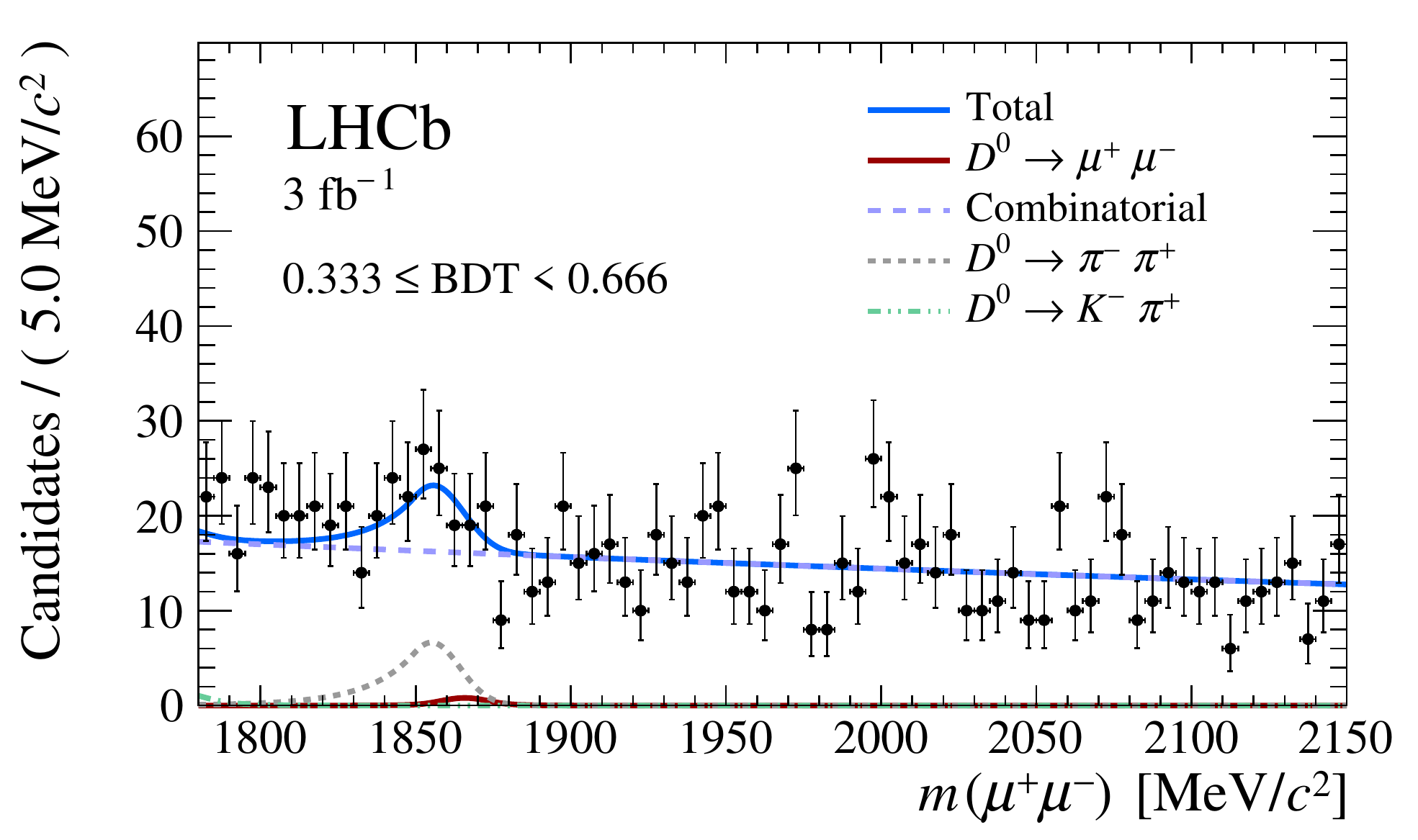}
\includegraphics[width = 0.5\textwidth]{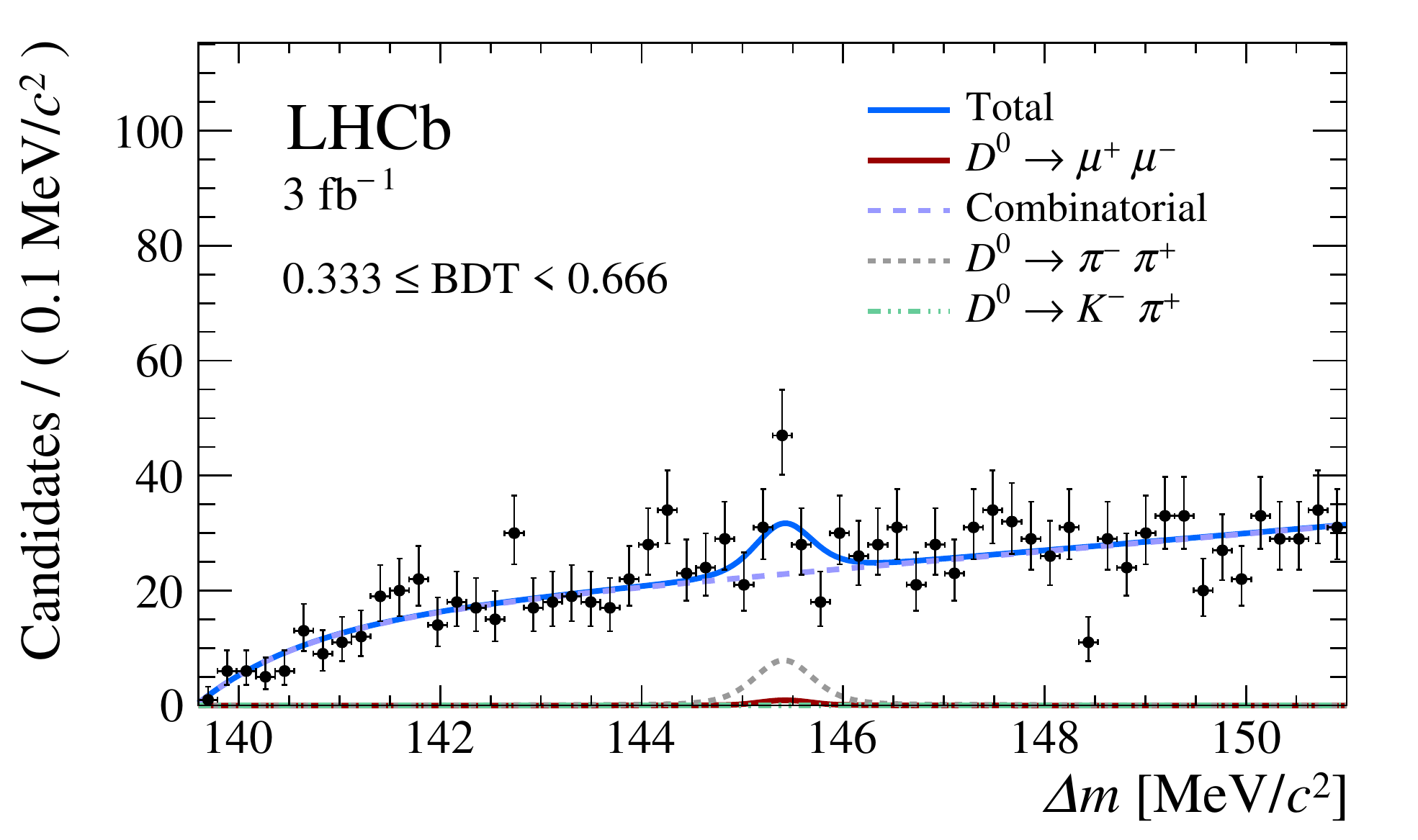}
\includegraphics[width = 0.5\textwidth]{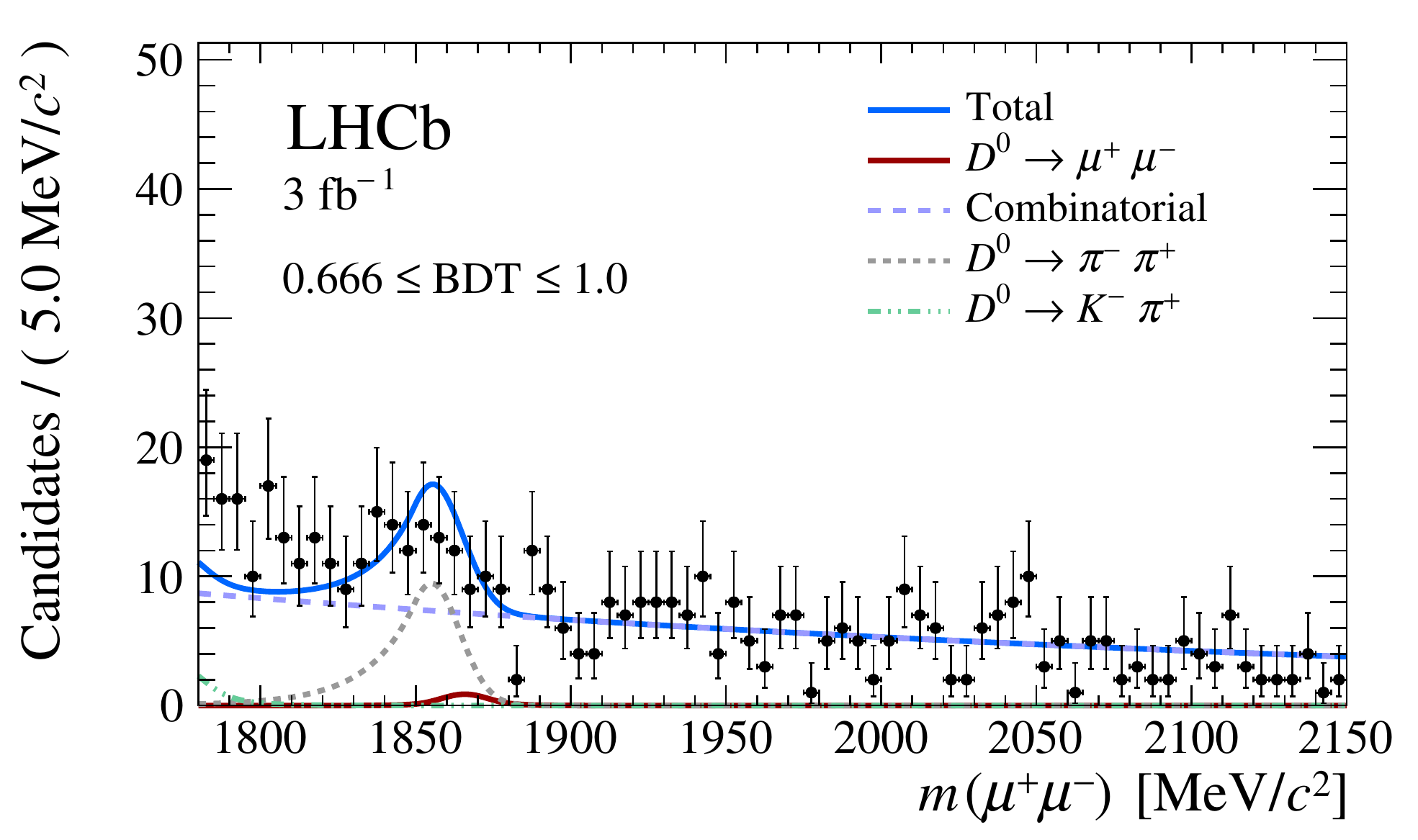}
\includegraphics[width = 0.5\textwidth]{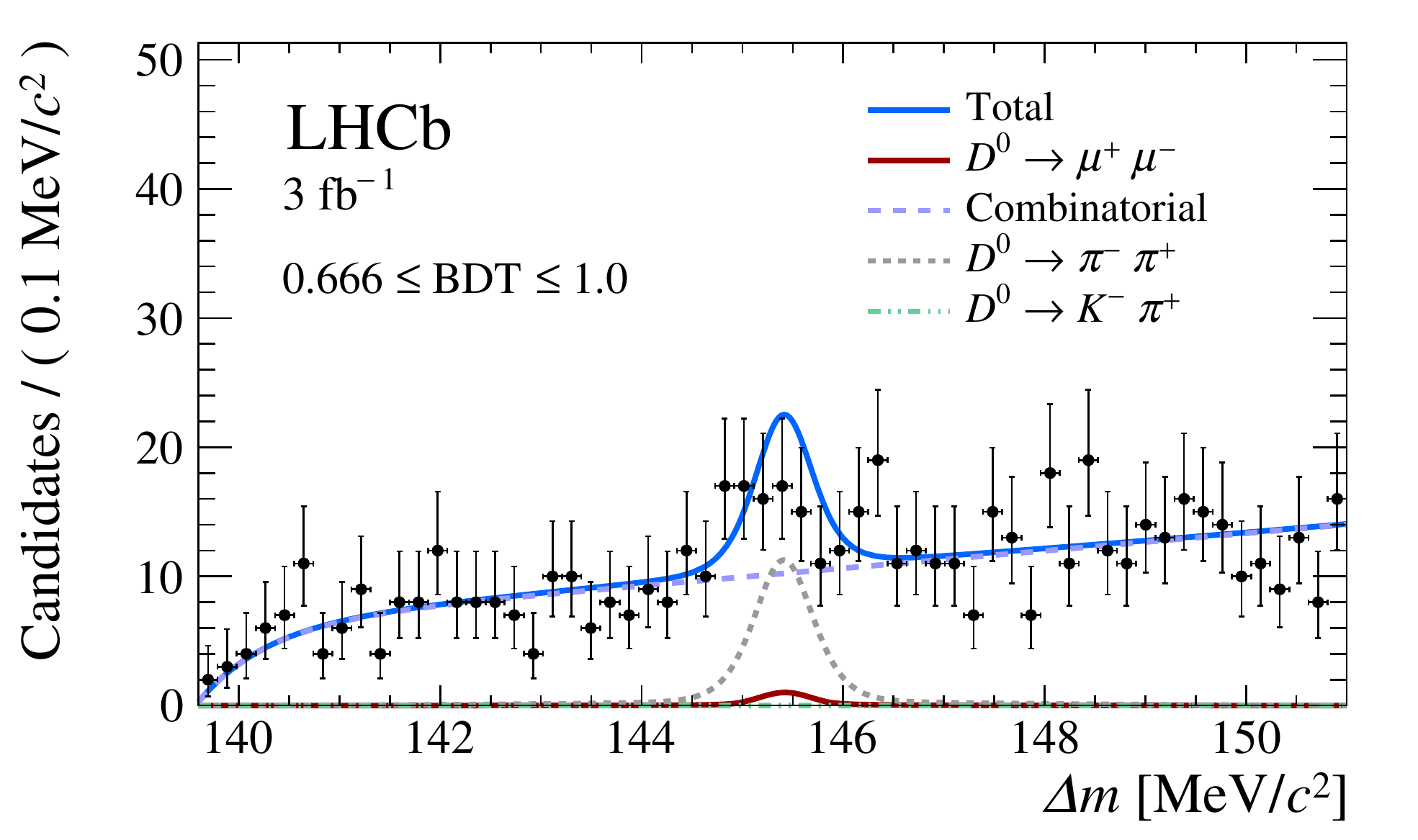}
 \caption{Distributions of (left) \mmumu and (right) \deltam  for the \dmumu  candidates for \runone data in, top to bottom, the three BDT intervals.
 The distributions are superimposed with the fit to data. 
  Each of the two distributions is in the signal region of the other variable, see text for details.
  Untagged and tagged decays are included in a single component for signal and \dpipi background.
 }\label{fig:signalplot_run1}
\end{figure}

\begin{figure}[btp]
\includegraphics[width = 0.5\textwidth]{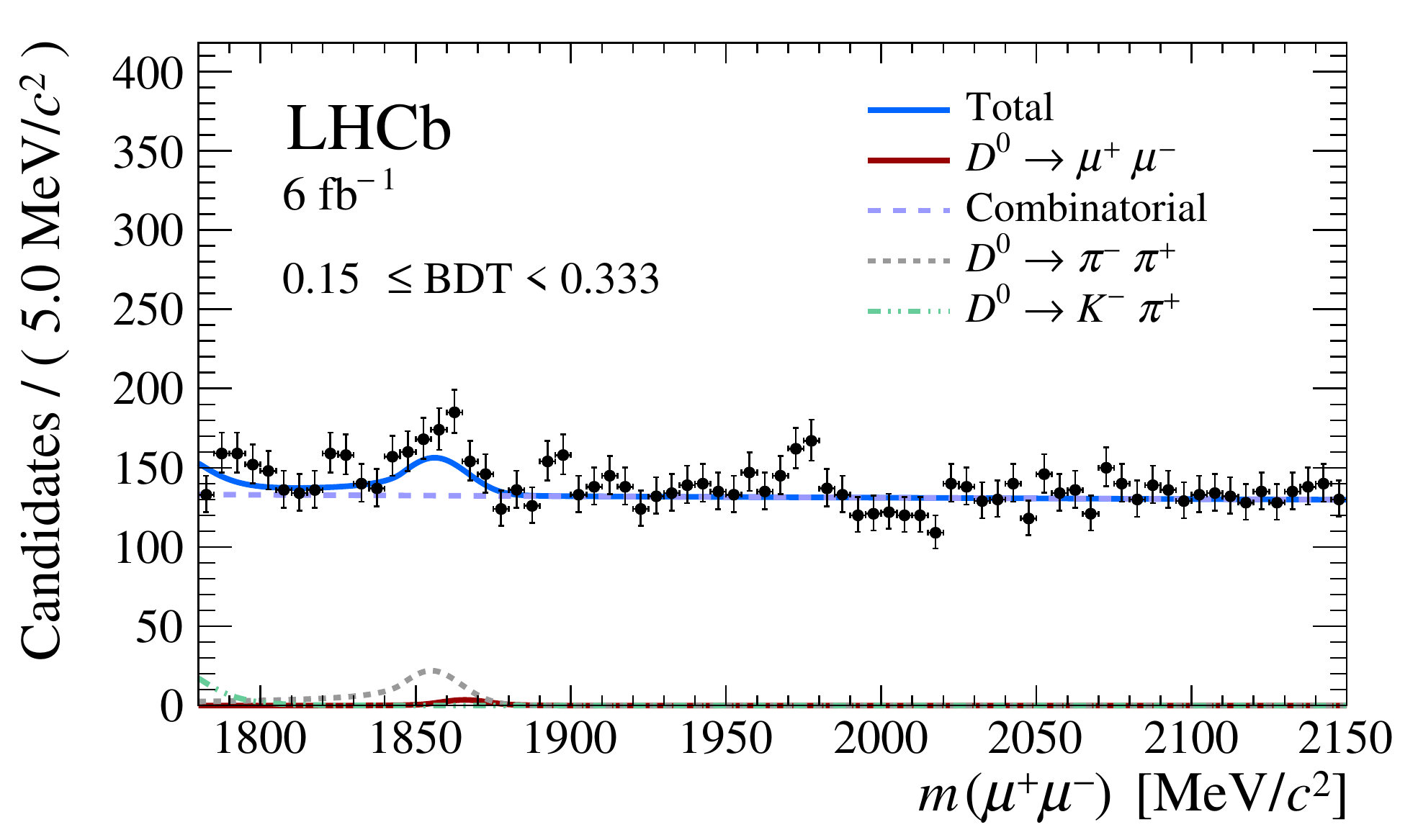}
\includegraphics[width = 0.5\textwidth]{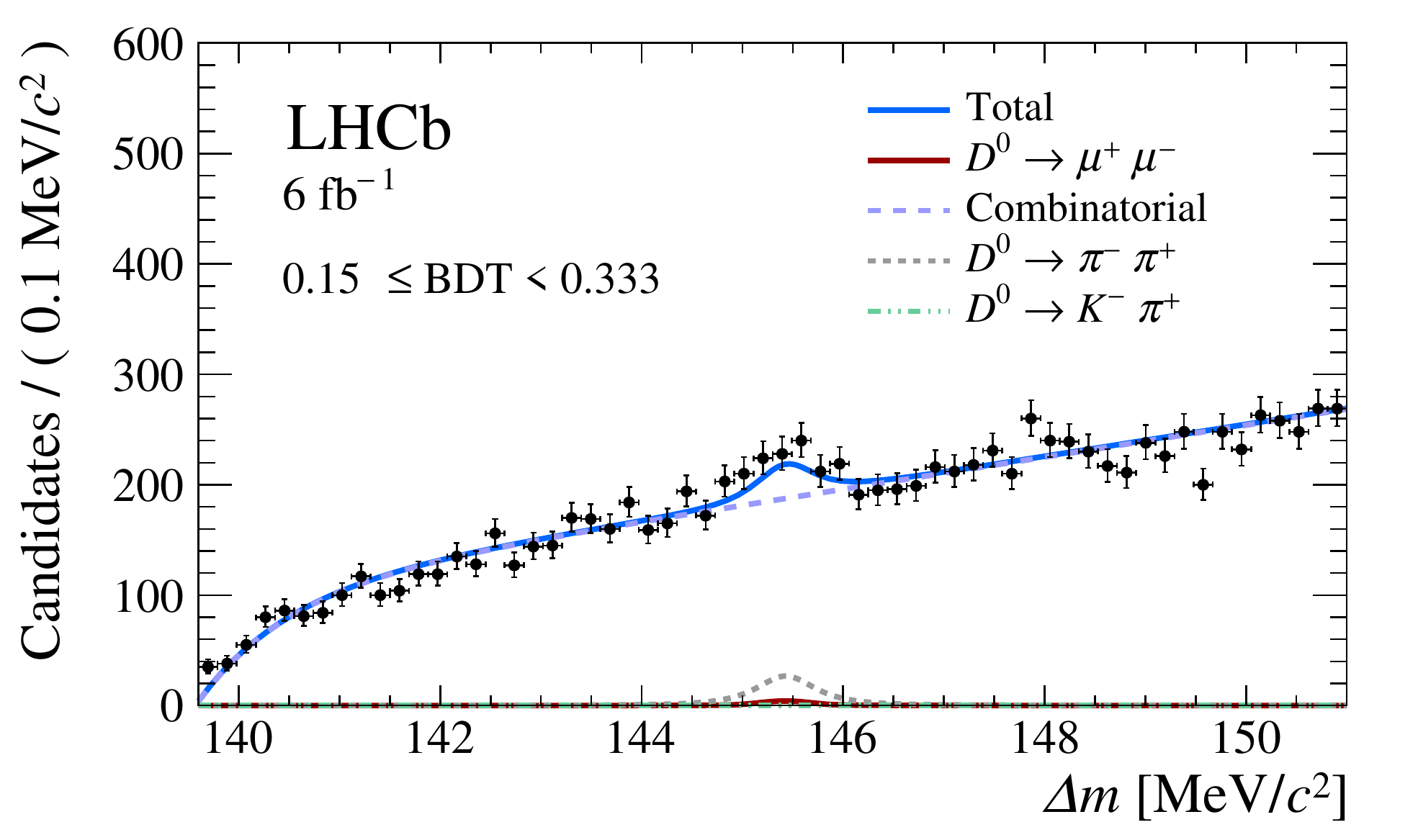}
\includegraphics[width = 0.5\textwidth]{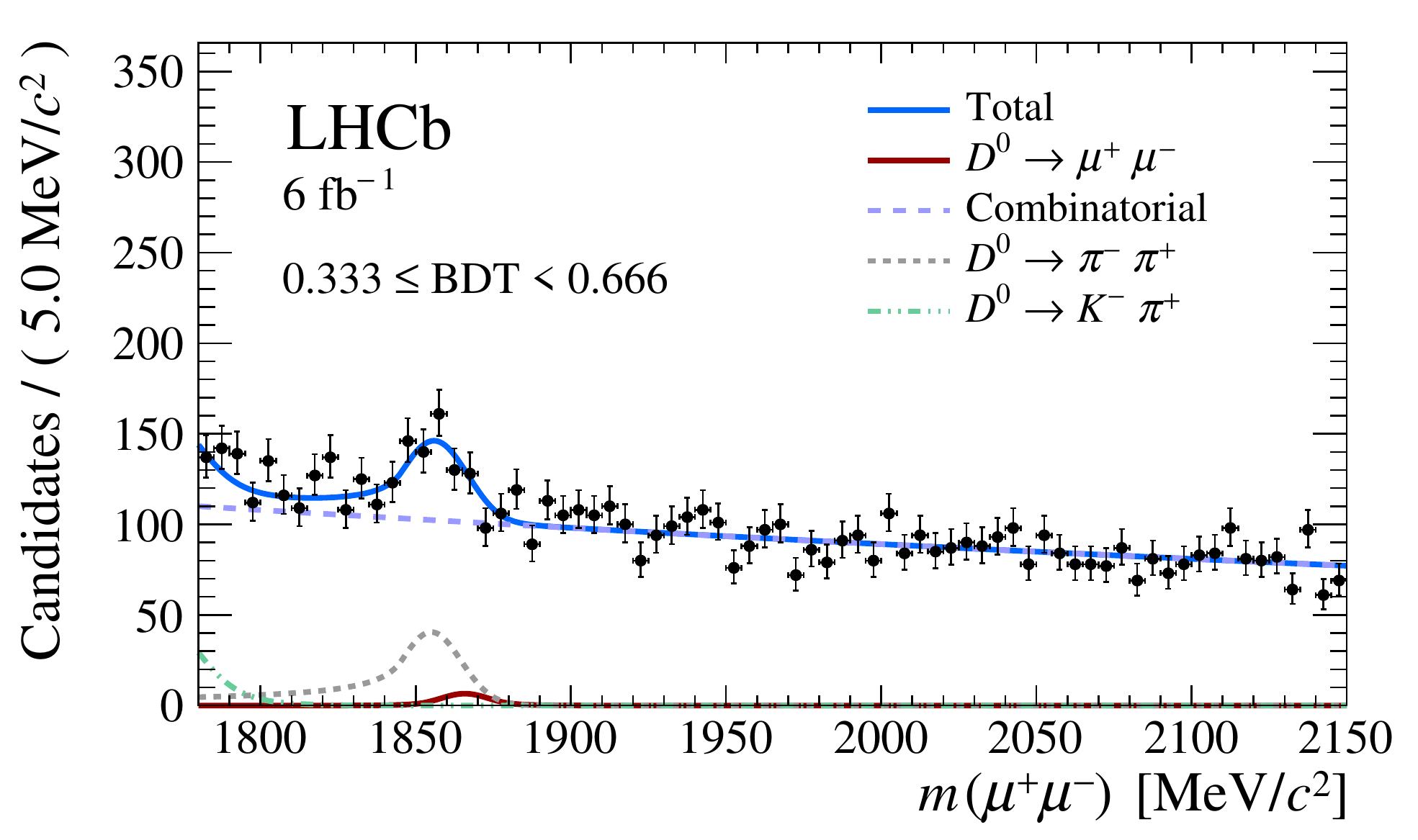}
\includegraphics[width = 0.5\textwidth]{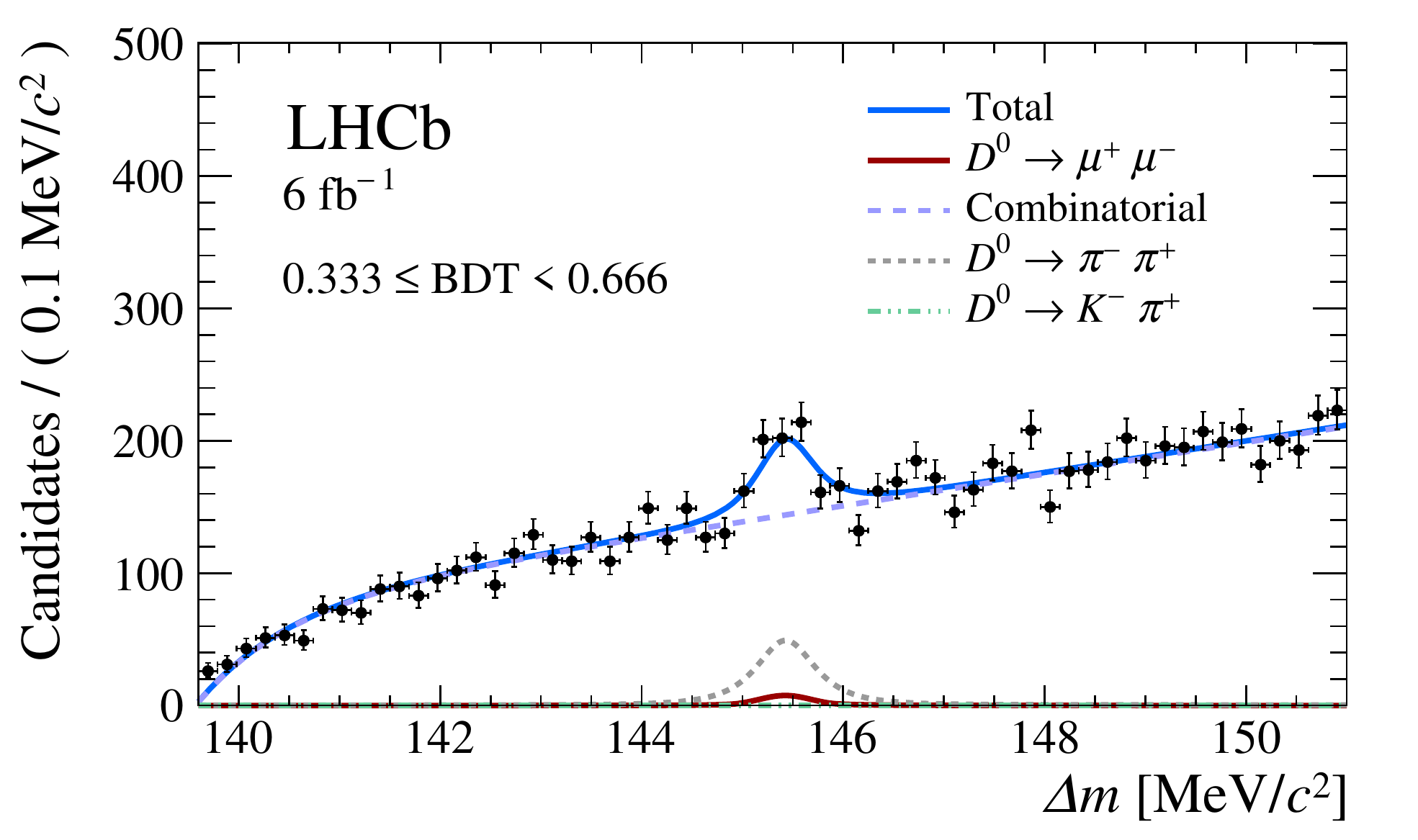}
\includegraphics[width = 0.5\textwidth]{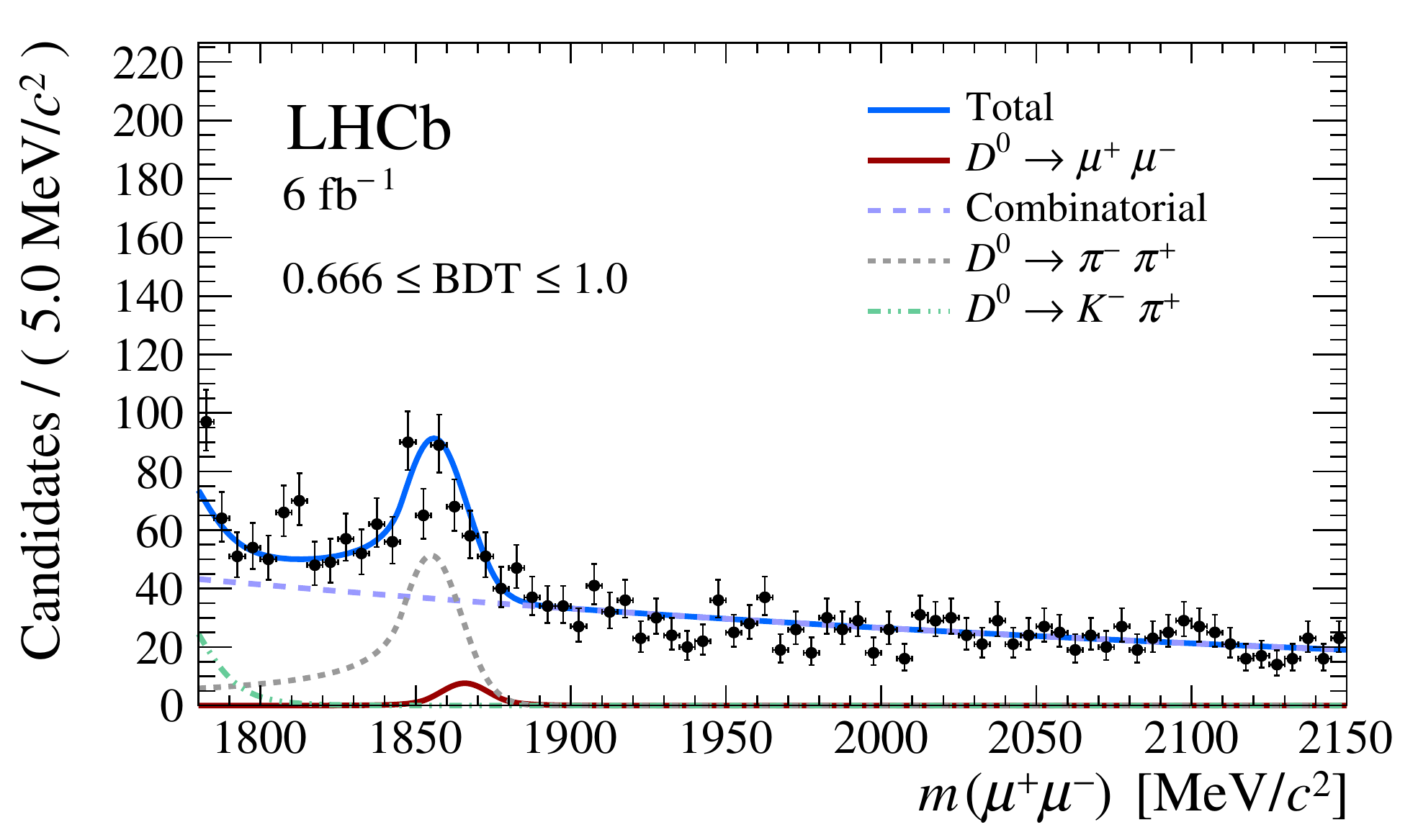}
\includegraphics[width = 0.5\textwidth]{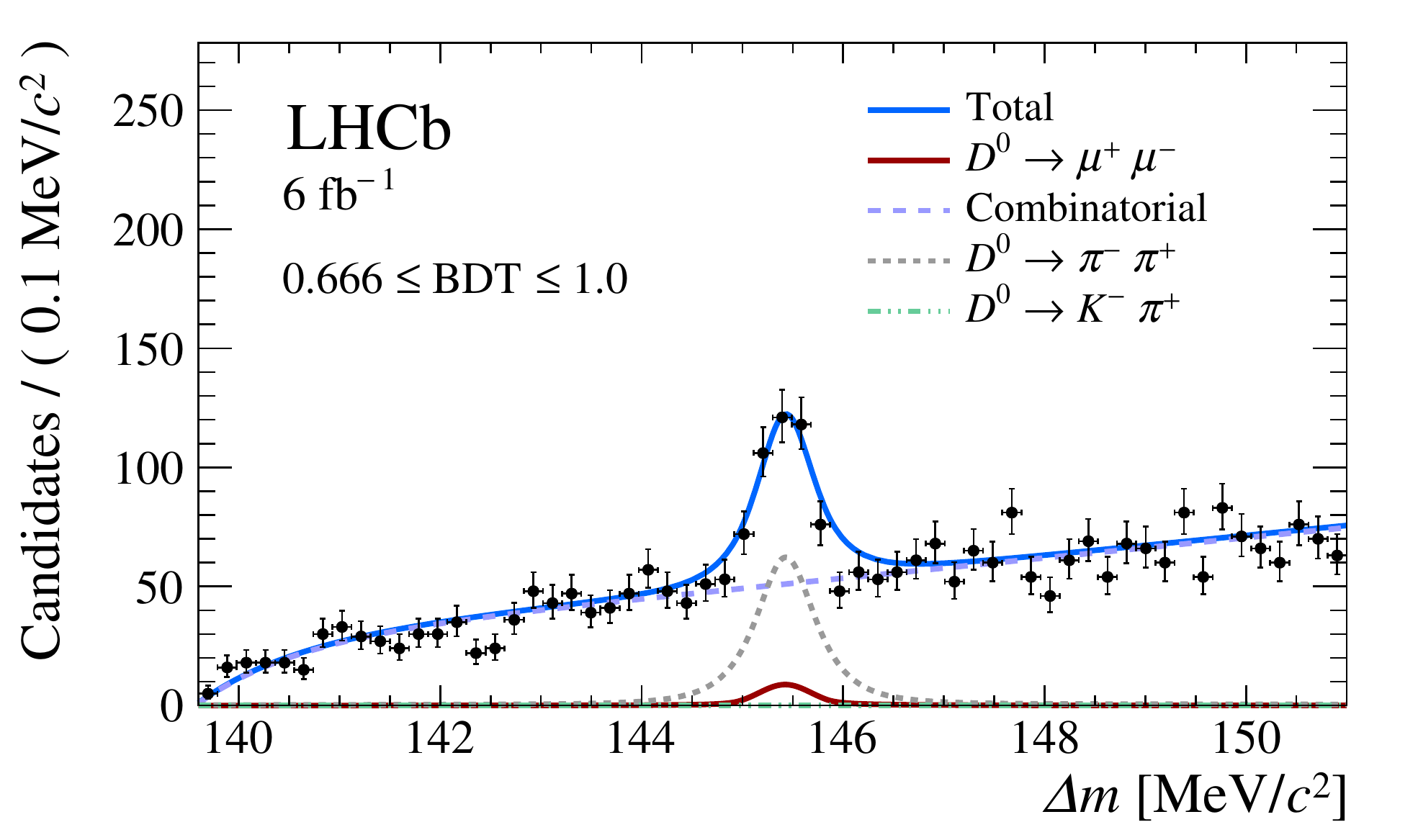}
\caption{Distributions of (left) \mmumu  and (right) \deltam  for the \dmumu  candidates for \runtwo data in, top to bottom, the three BDT bins.
 The distributions are superimposed with the fit to data. 
  Each of the two distributions is in the signal region of the other variable, see text for details.    
  Untagged and tagged decays are included in a single component for signal and \dpipi background.
 }\label{fig:signalplot_run2}
\end{figure}

\begin{figure}[bp]
\includegraphics[width = 0.5\textwidth]{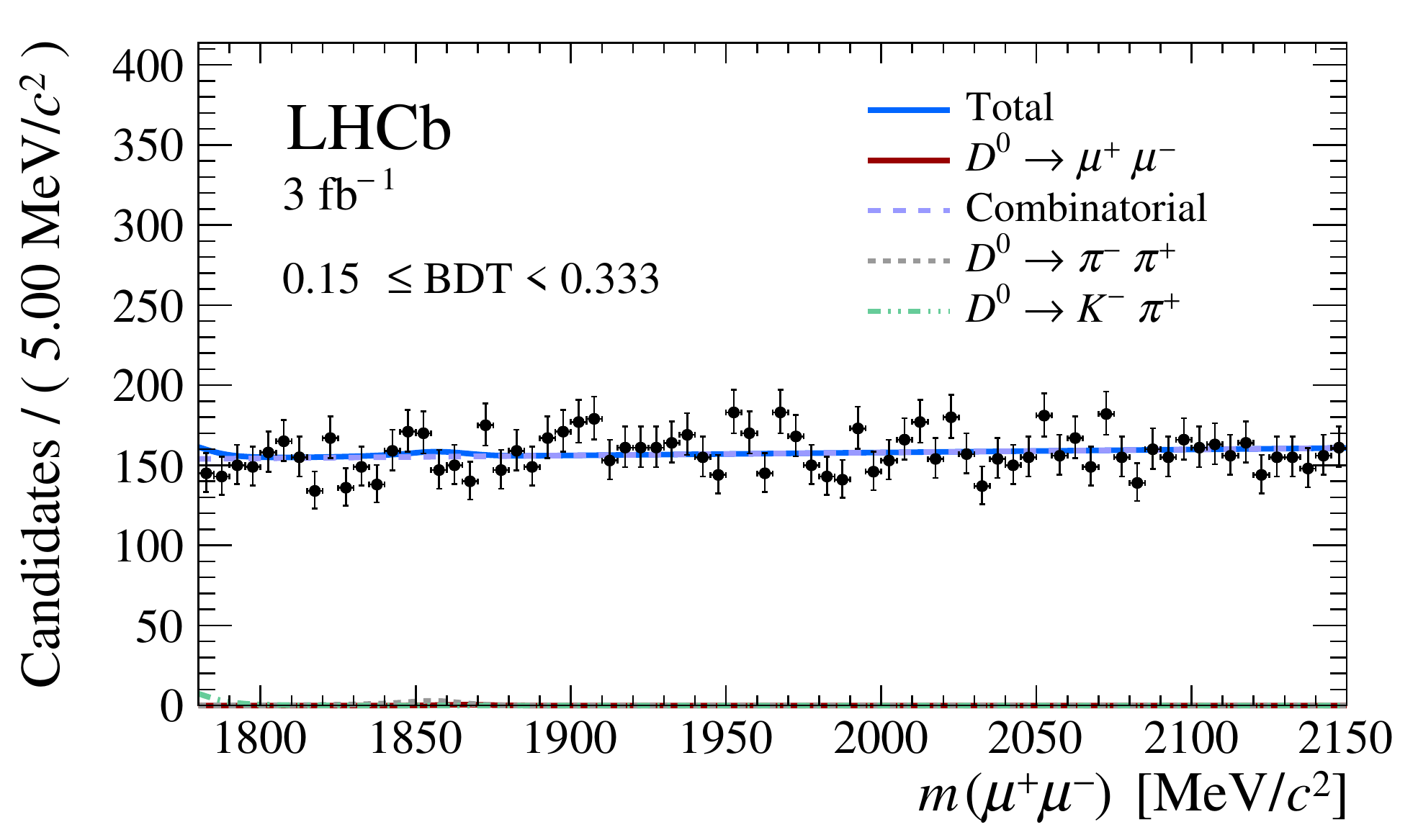}
\includegraphics[width = 0.5\textwidth]{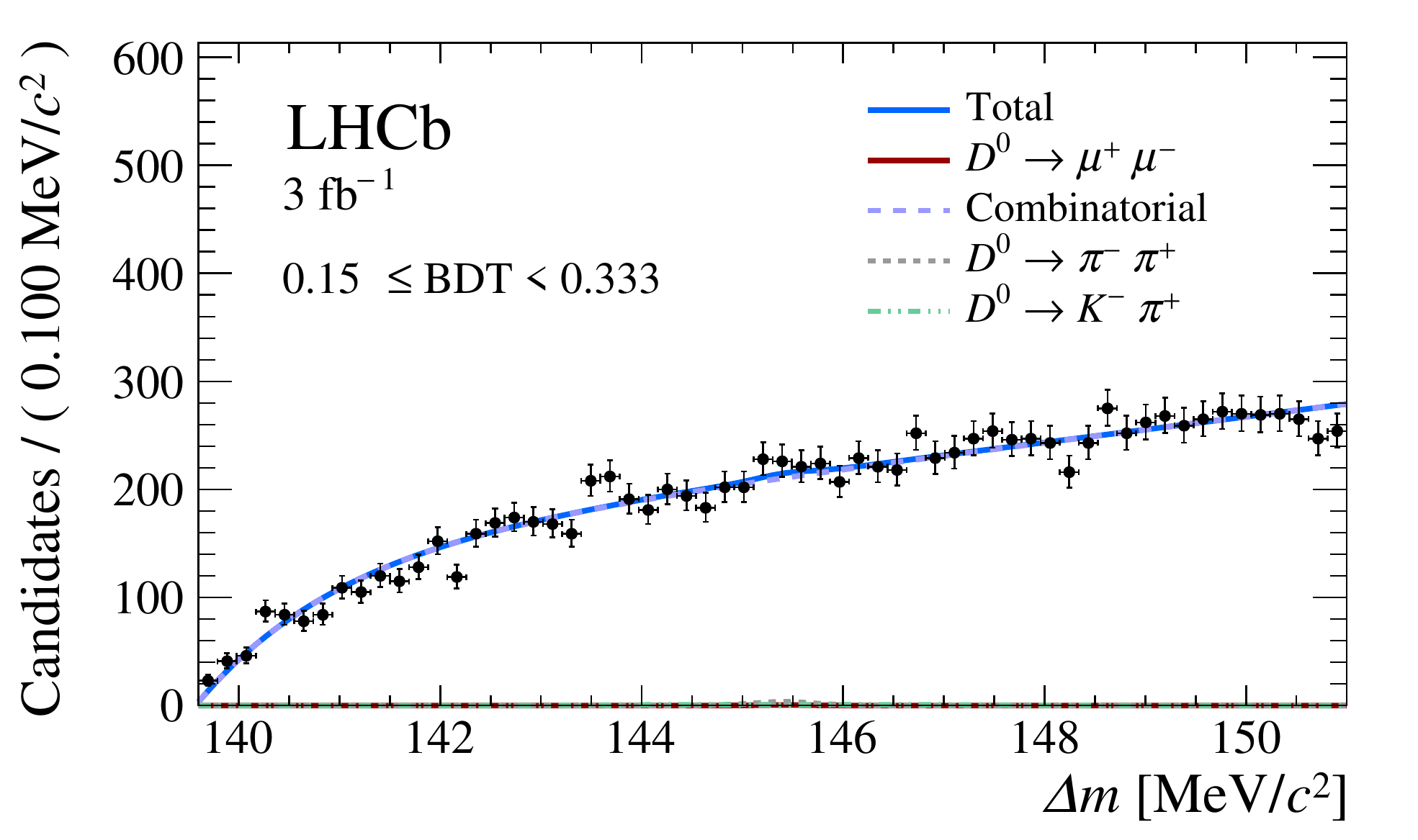}
\includegraphics[width = 0.5\textwidth]{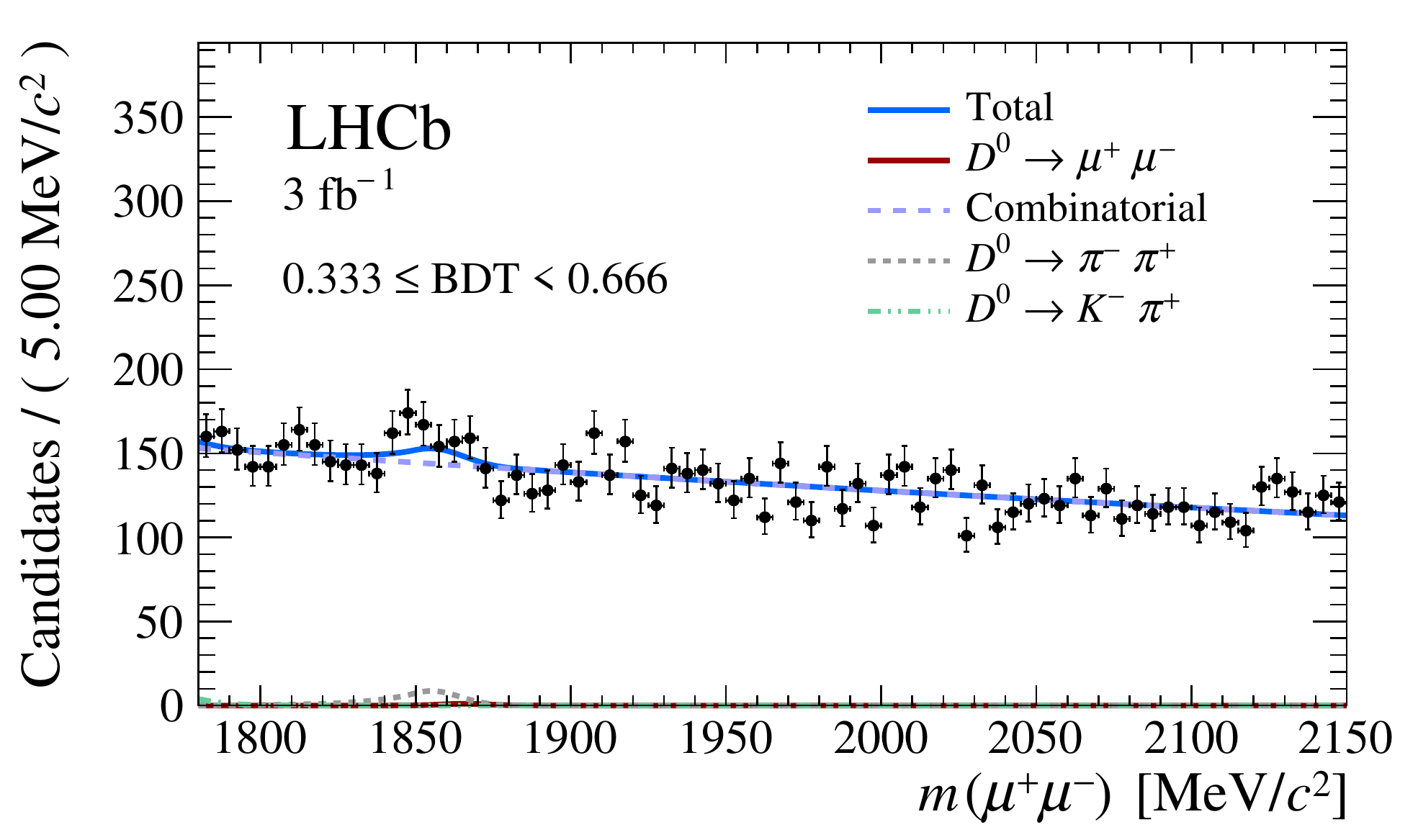}
\includegraphics[width = 0.5\textwidth]{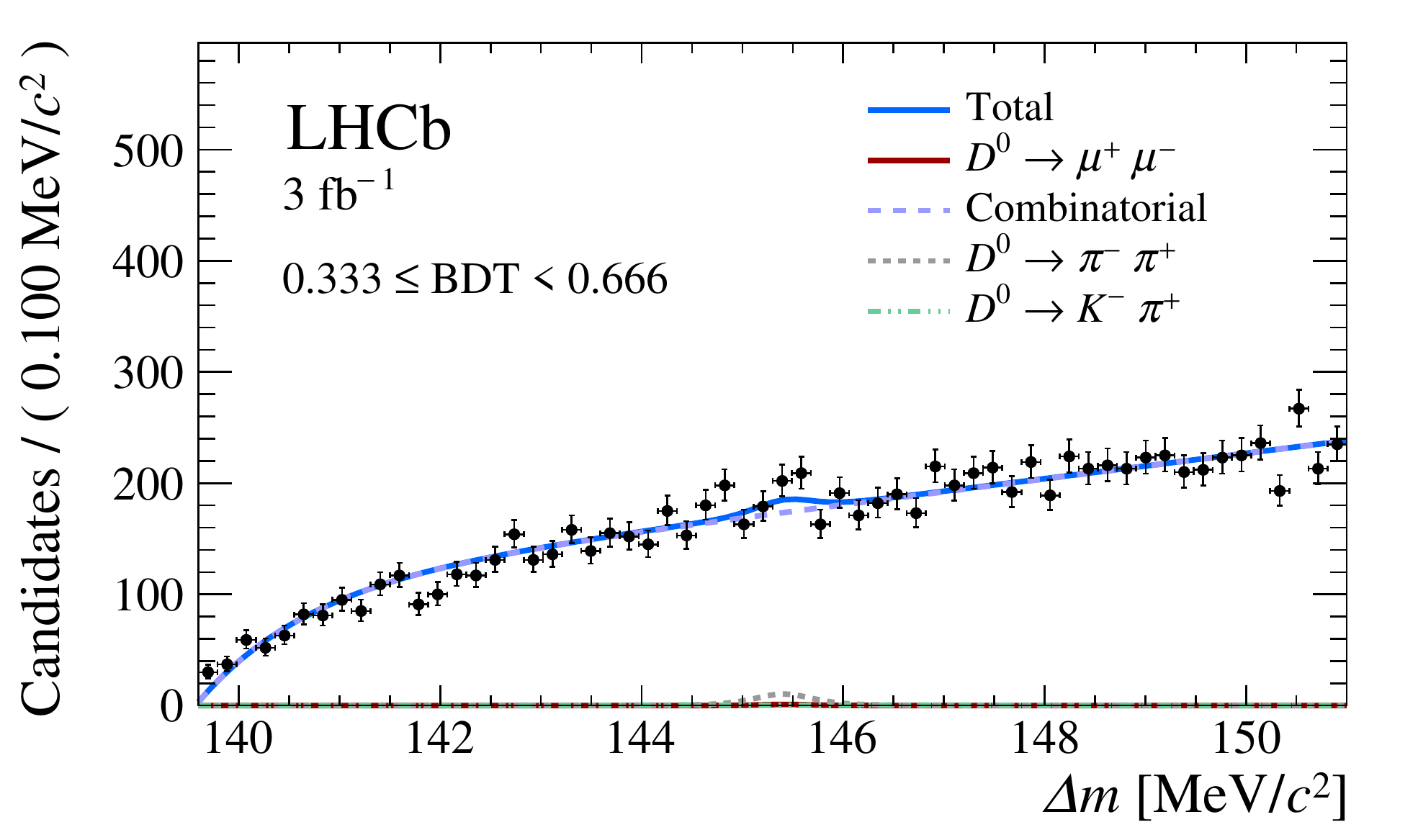}
\includegraphics[width = 0.5\textwidth]{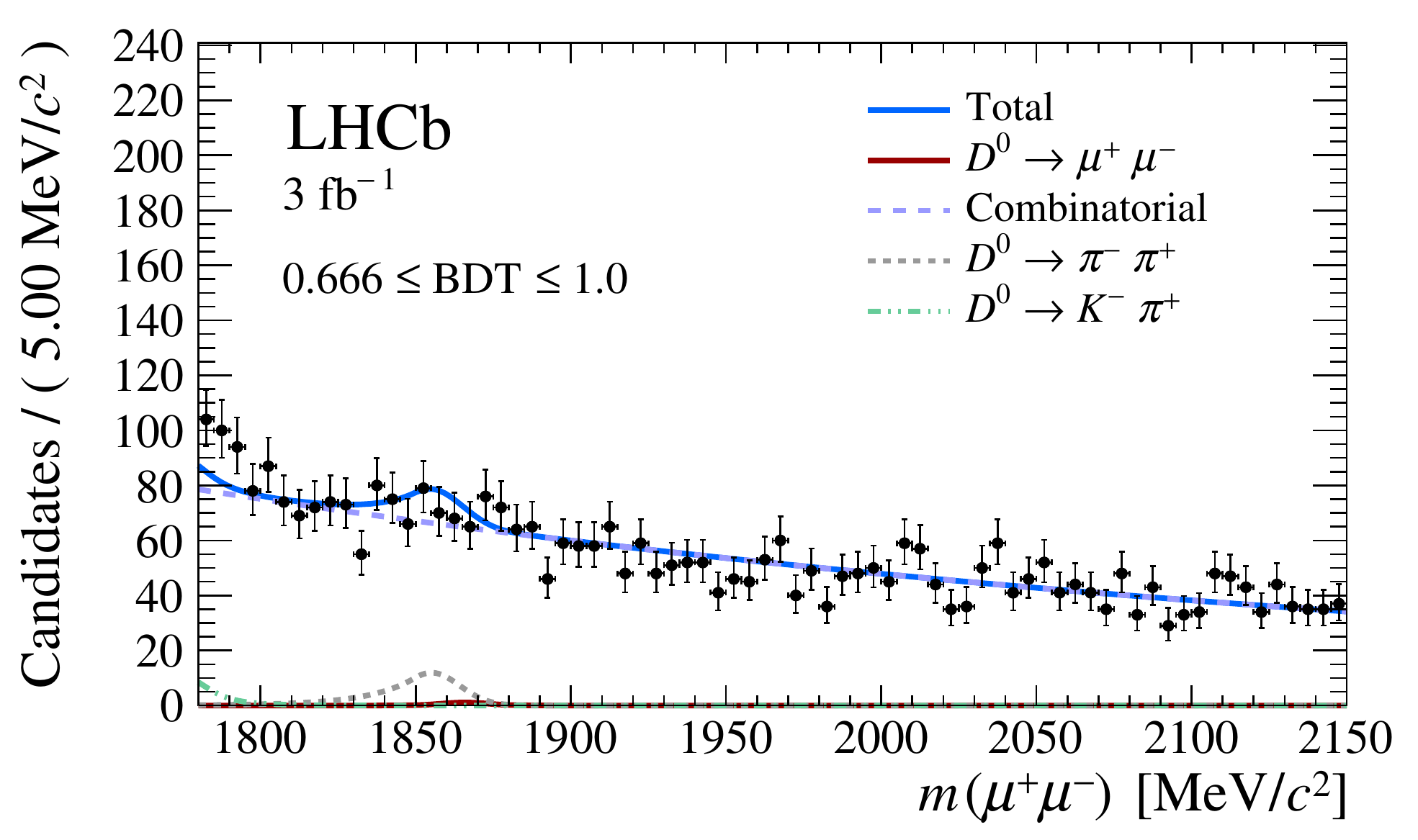}
\includegraphics[width = 0.5\textwidth]{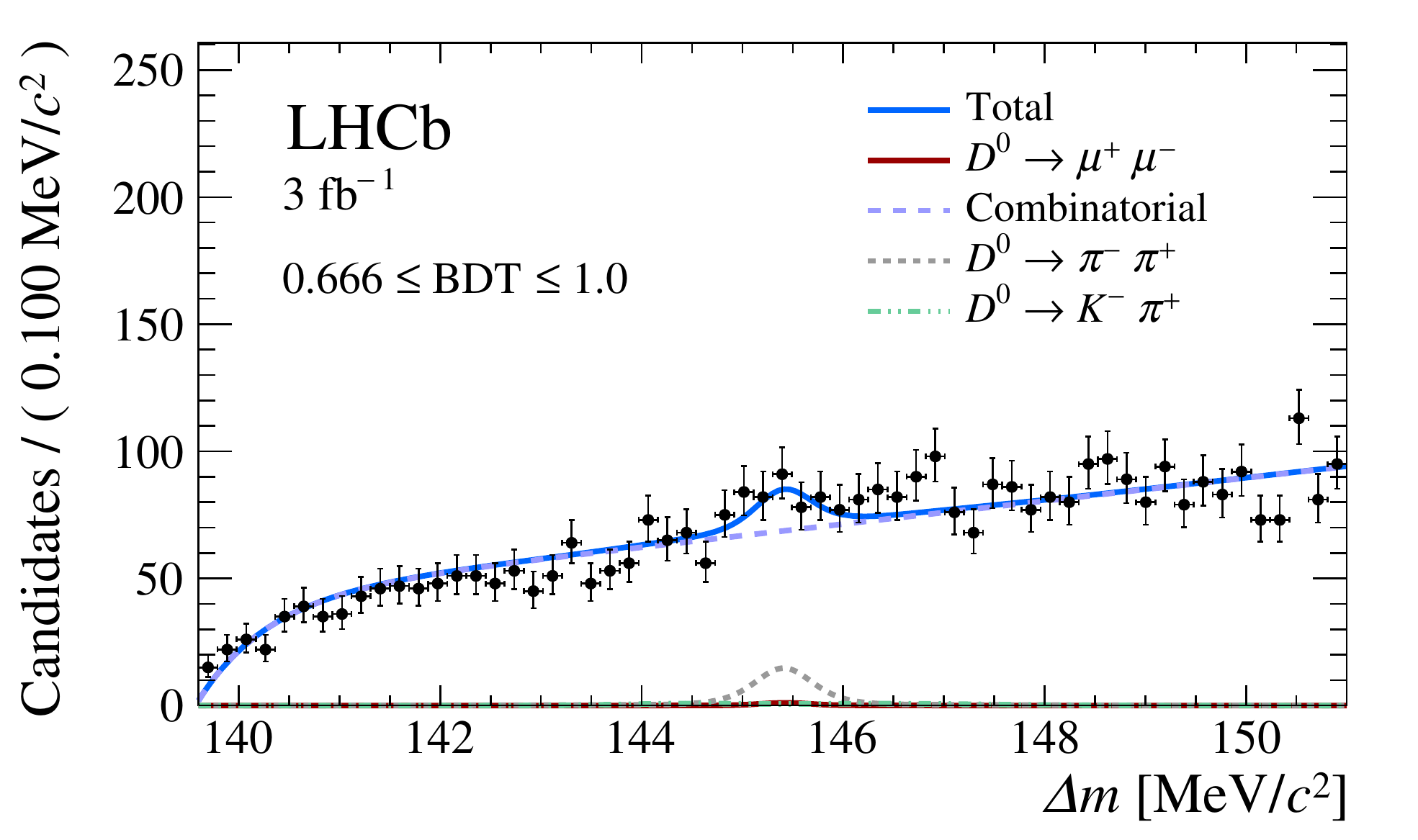}
 \caption{Distributions of (left) \mmumu  and (right) \deltam  for the \dmumu  candidates for Run 1 data in, top to bottom, the three BDT bins.
 The distributions are superimposed with the fit to data. 
  Untagged and tagged decays are included in a single component for signal and \dpipi background. Unlike the correspondent figures in the main body of the Letter, here all events are shown.
 }\label{fig:signalplot_uncut_run1}
\end{figure}

\begin{figure}
\includegraphics[width = 0.5\textwidth]{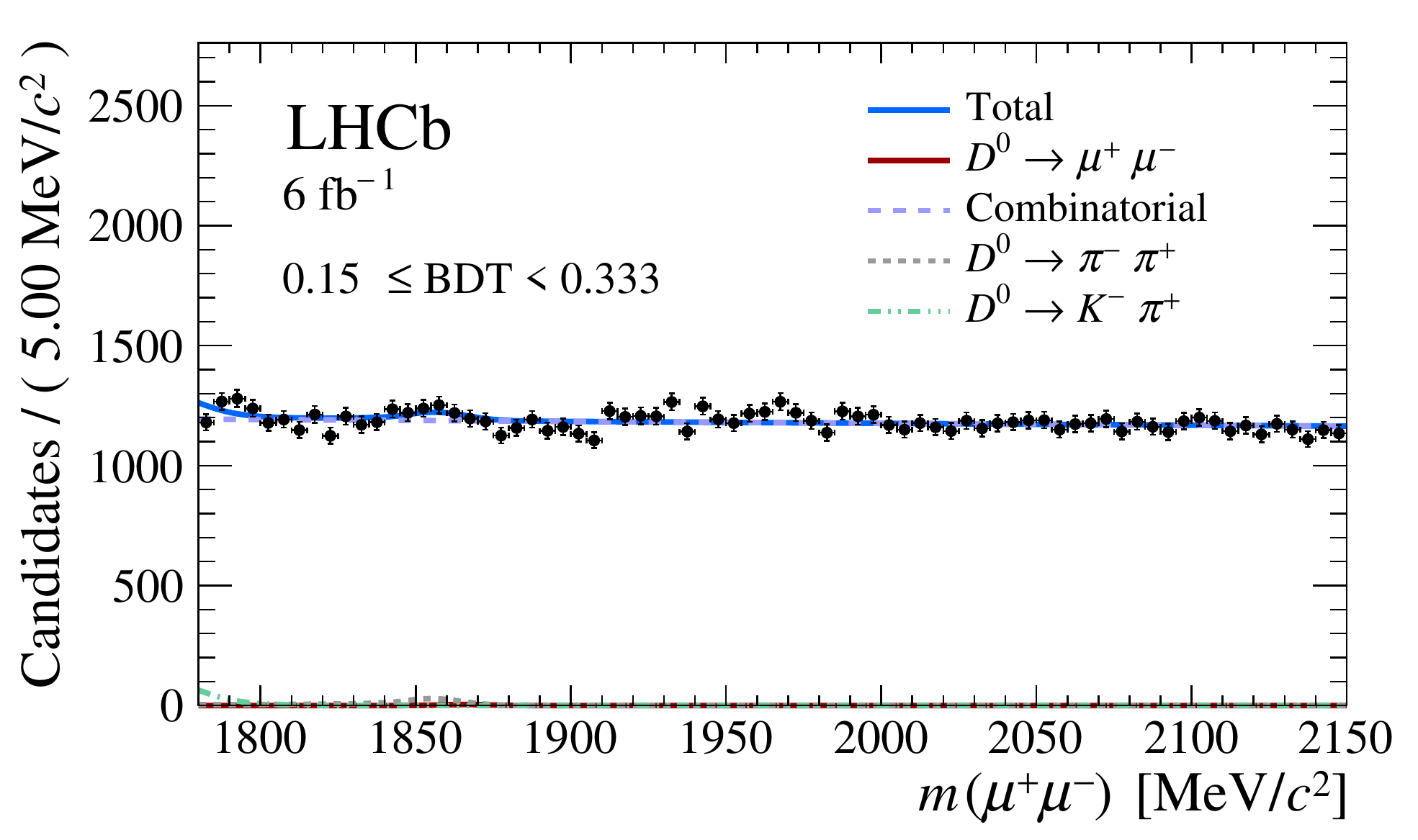}
\includegraphics[width = 0.5\textwidth]{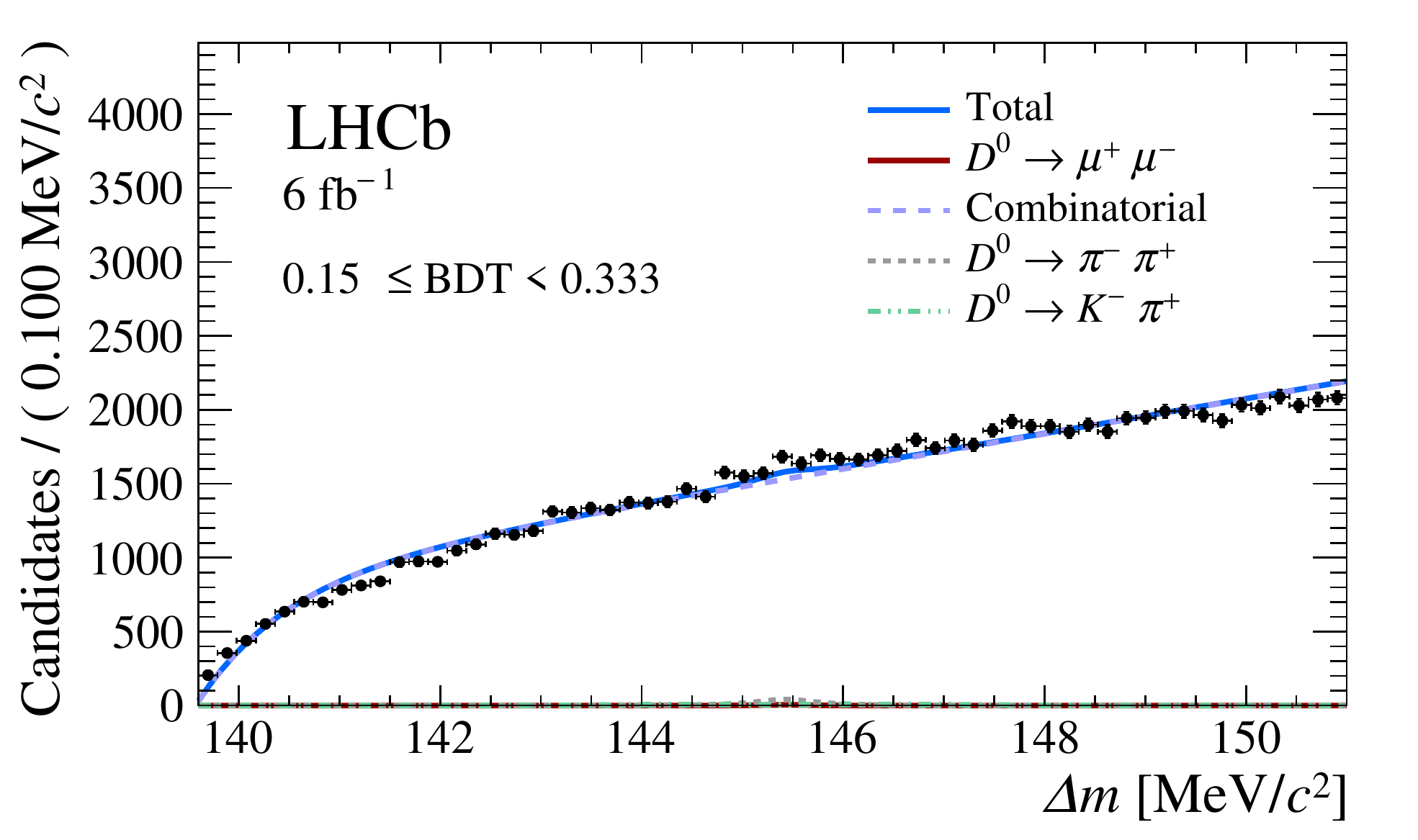}
\includegraphics[width = 0.5\textwidth]{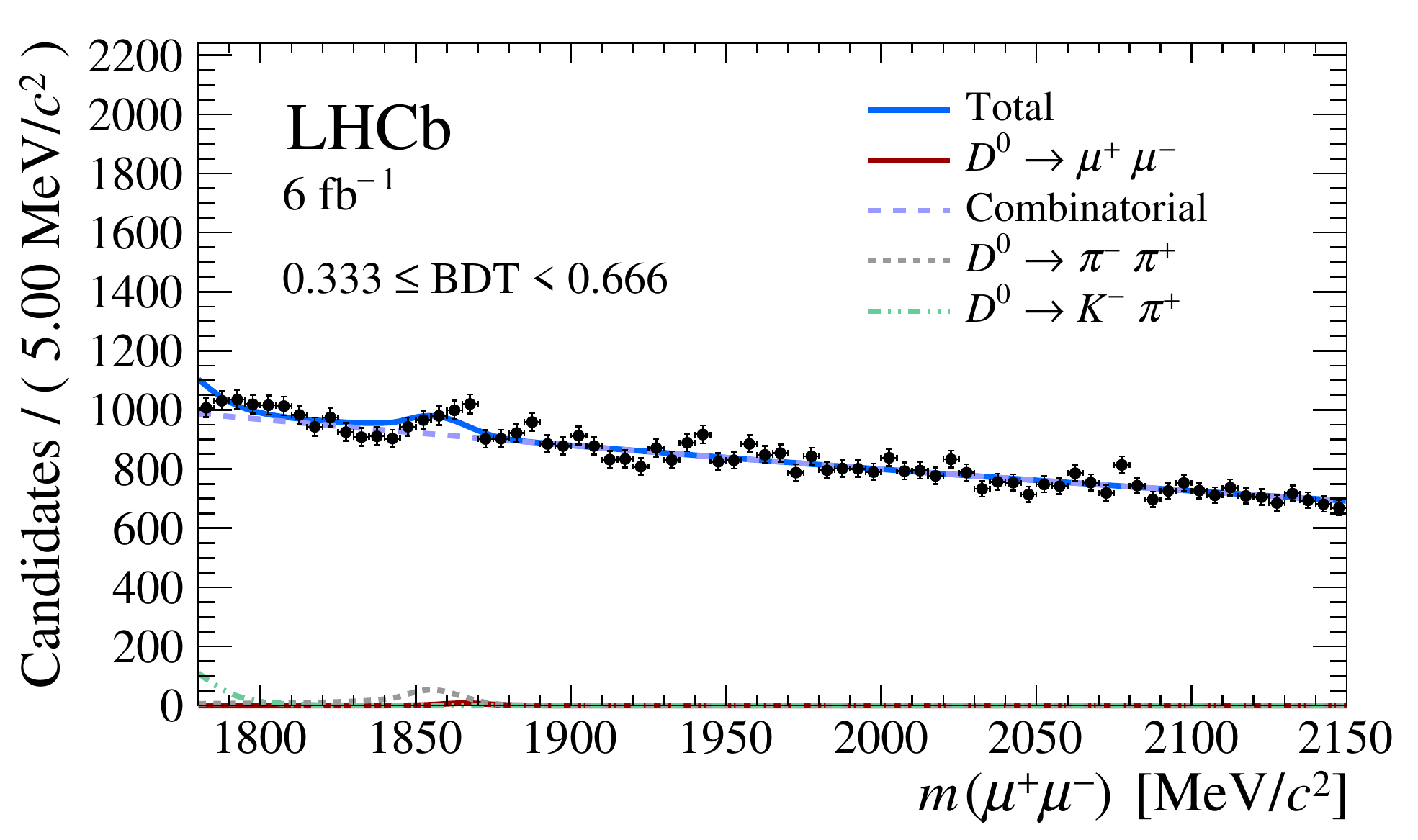}
\includegraphics[width = 0.5\textwidth]{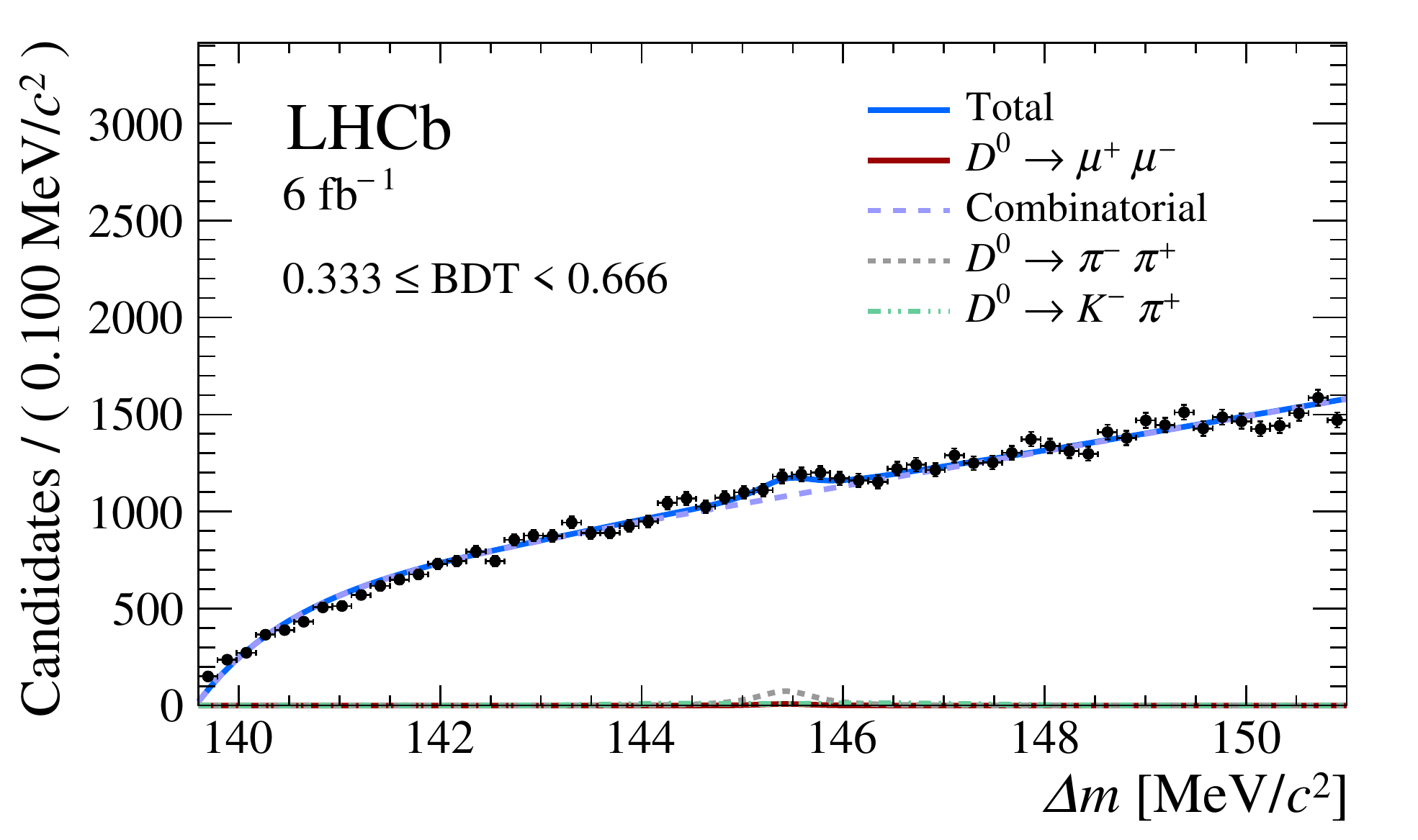}
\includegraphics[width = 0.5\textwidth]{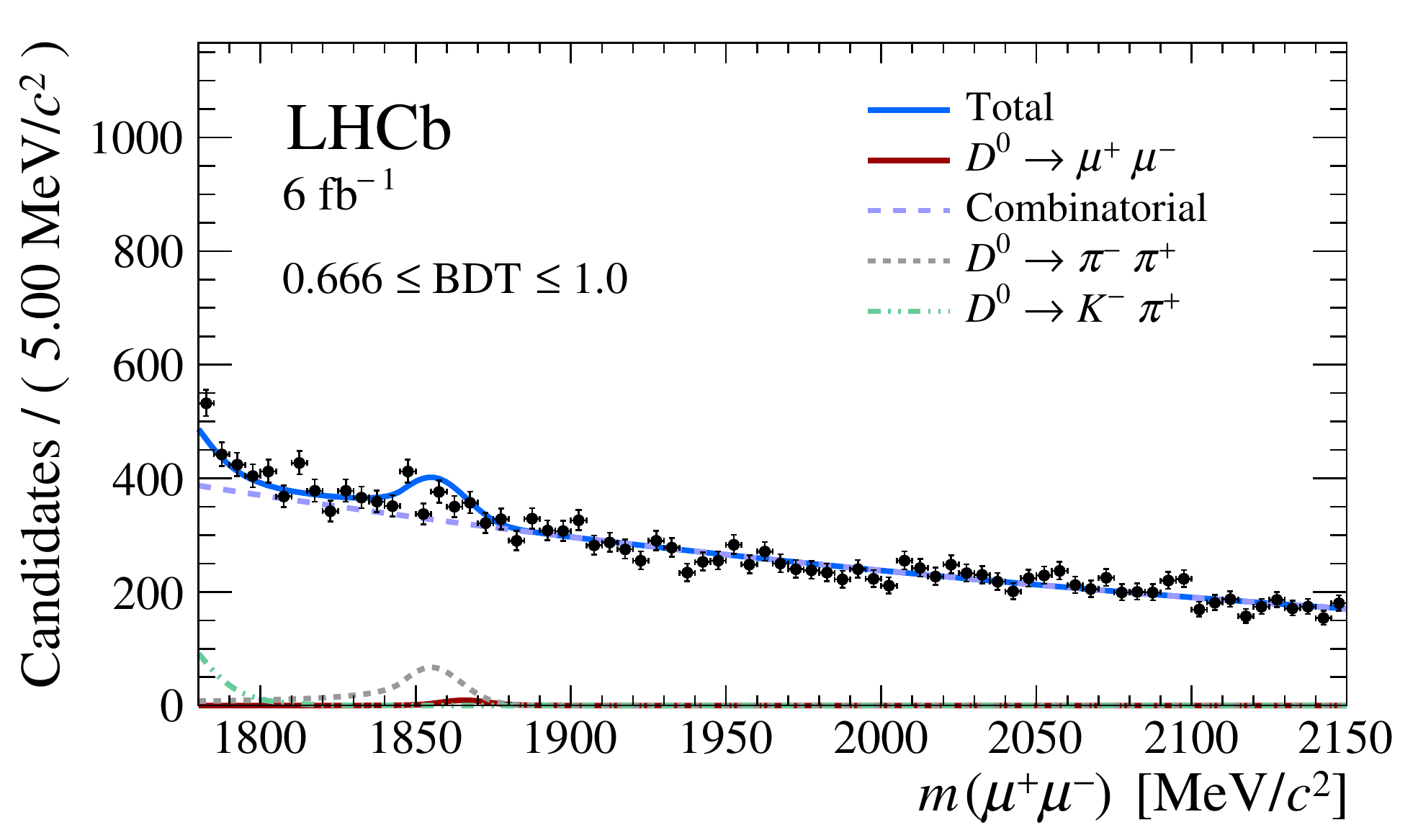}
\includegraphics[width = 0.5\textwidth]{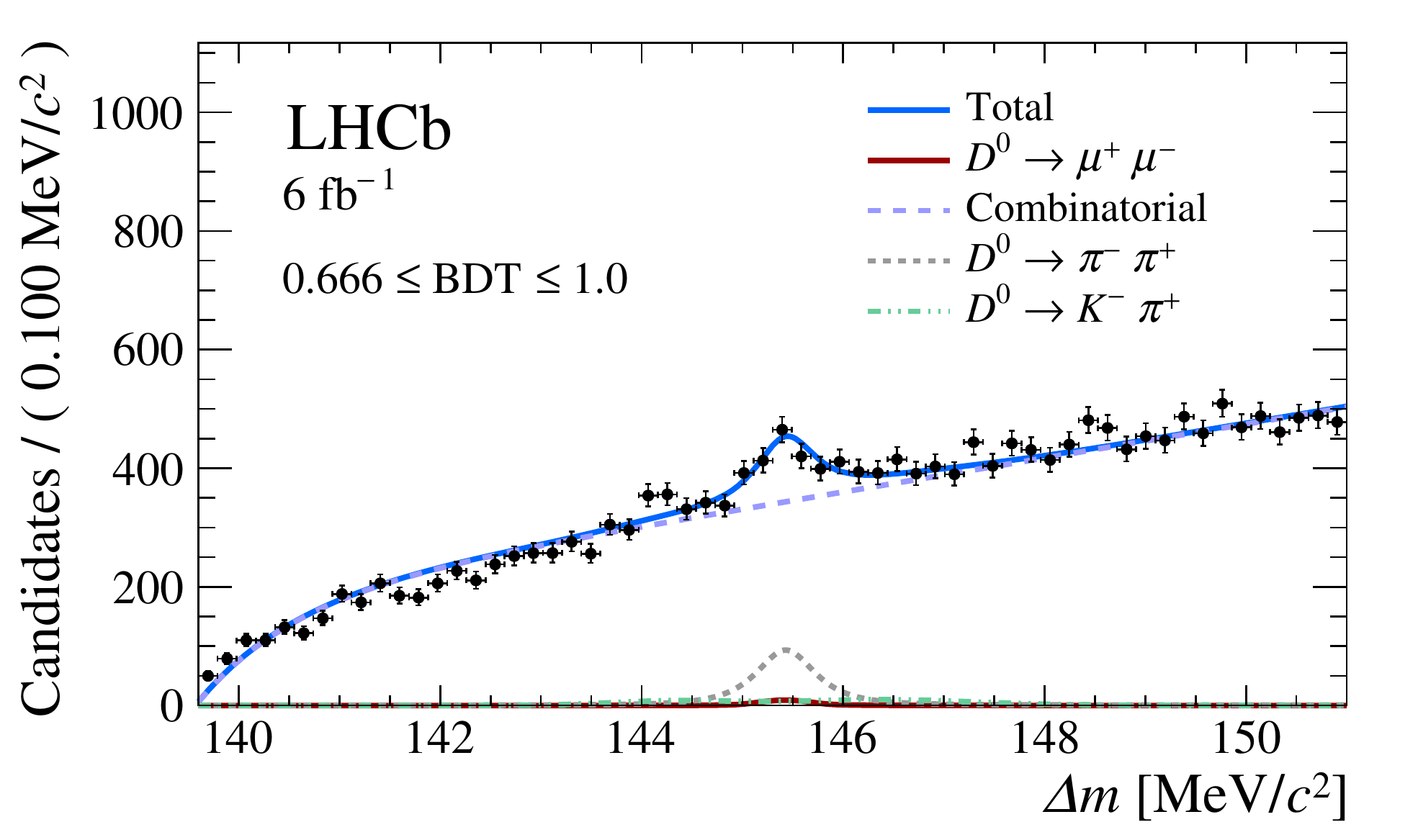}
\caption{Distributions of (left) \mmumu  and (right) \deltam  for the \dmumu  candidates for Run 2 data in, top to bottom, the three BDT bins.
 The distributions are superimposed with the fit to data. 
  Untagged and tagged decays are included in a single component for signal and \dpipi background. Unlike the correspondent figures in the main body of the Letter, here all events are shown.
 }\label{fig:signalplot_uncut_run2}
\end{figure}

\begin{figure}
\includegraphics[width = 0.5\textwidth]{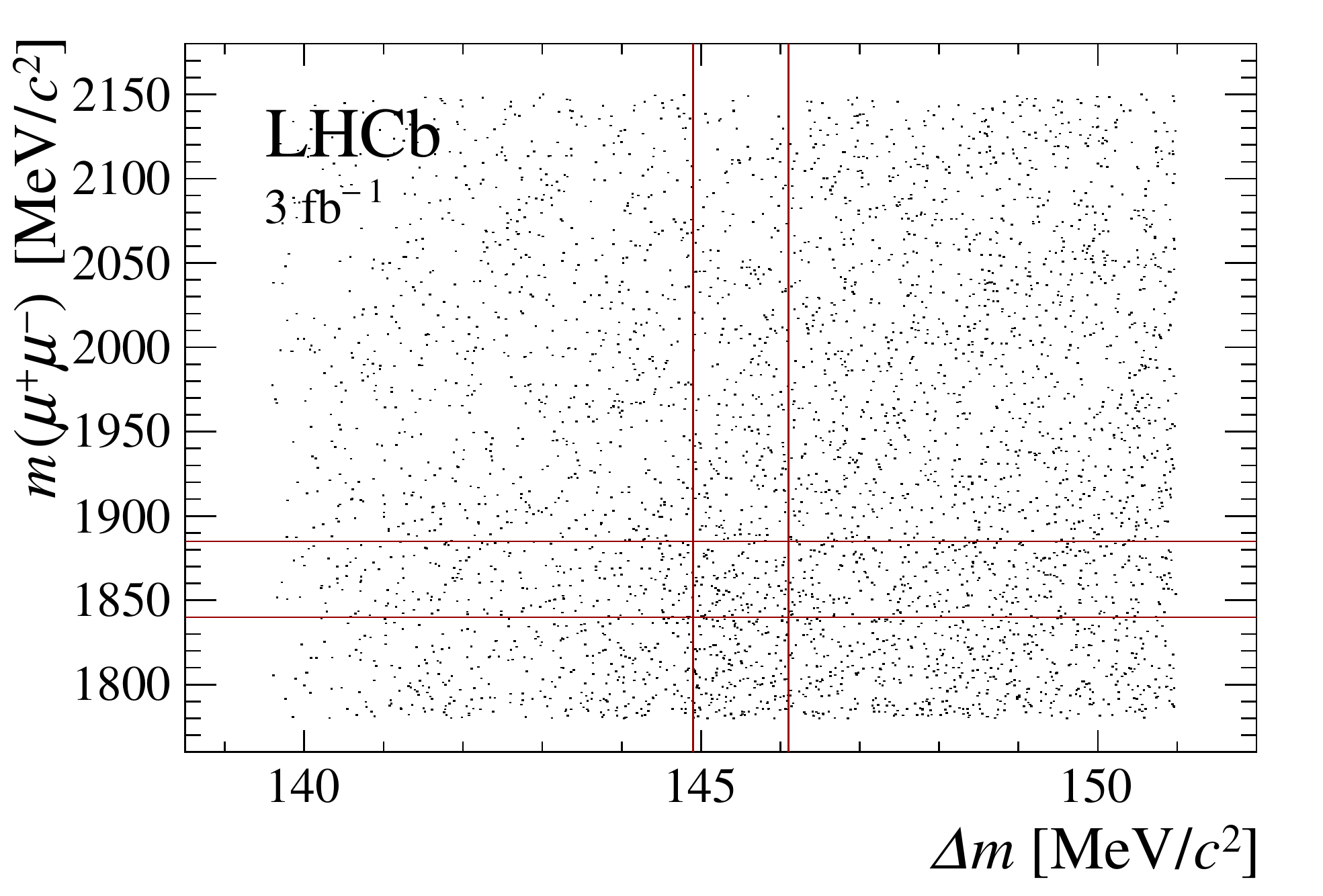}
\includegraphics[width = 0.5\textwidth]{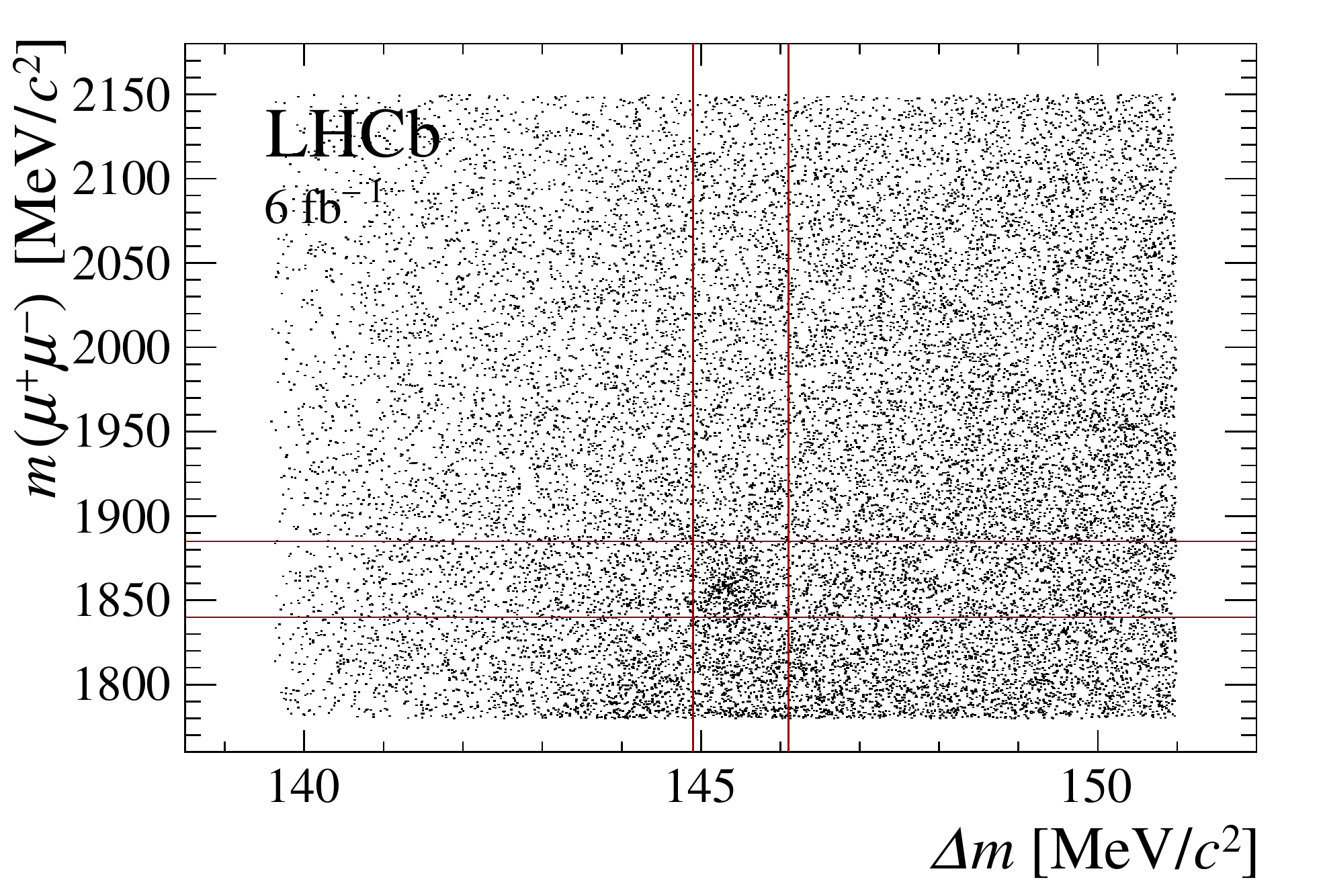}
\caption{Two-dimensional distribution of \md versus \deltam for the most sensitive BDT interval in the (left) Run~1 and (right) Run~2 data.  Red lines represent the signal region of each variable as defined in the text. }\label{fig:2dplot}
\end{figure}

\begin{figure}
\includegraphics[width = 0.5\textwidth]{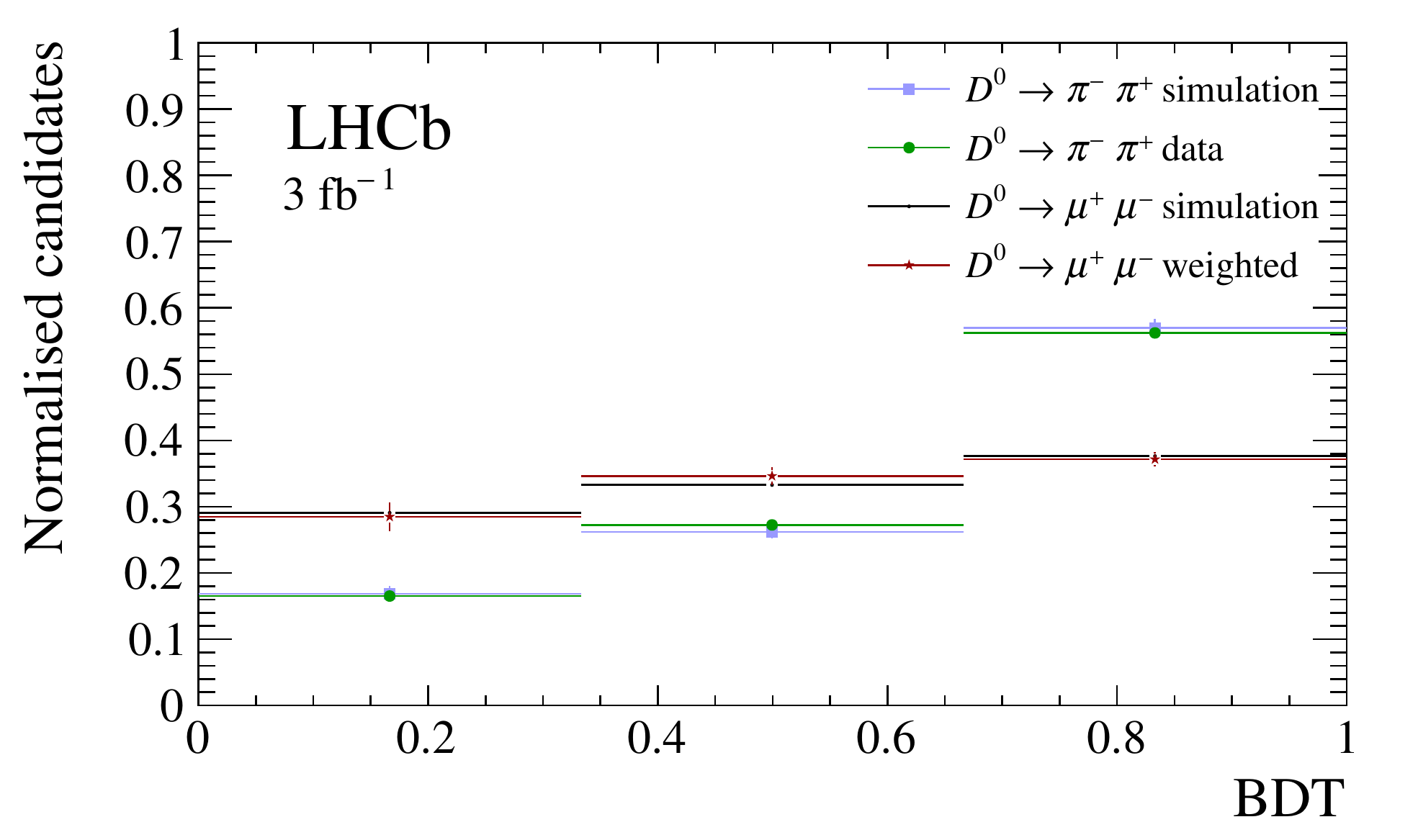}
\includegraphics[width = 0.5\textwidth]{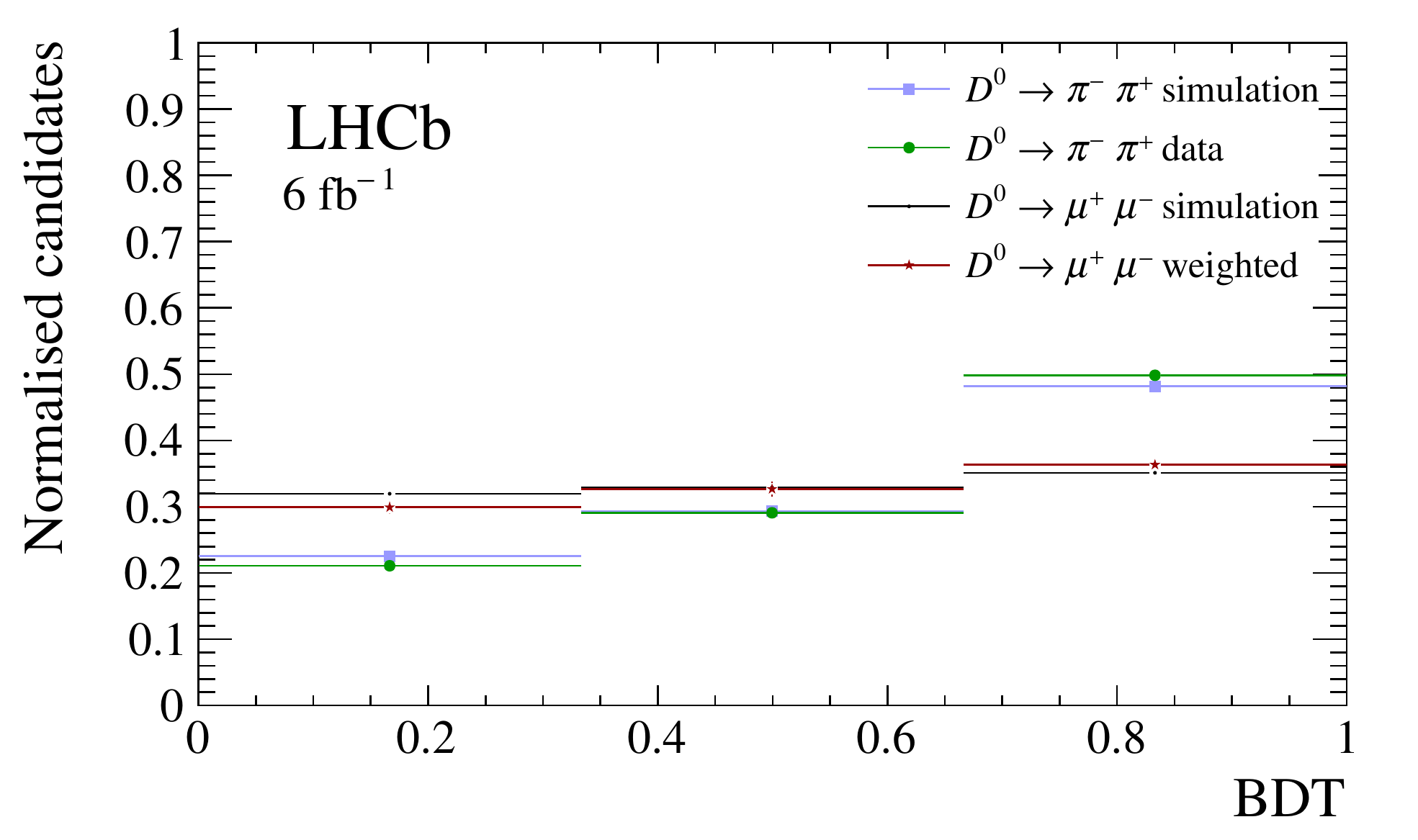}
\caption{Distribution of the BDT output. The \dpipi candidates in data and simulation are shown and 
used to weight the simulated \dmumu decays distribution used for the search.
 }\label{fig:bdtcalib}
\end{figure}

\begin{figure}
\centering
\includegraphics[width = 0.5\textwidth]{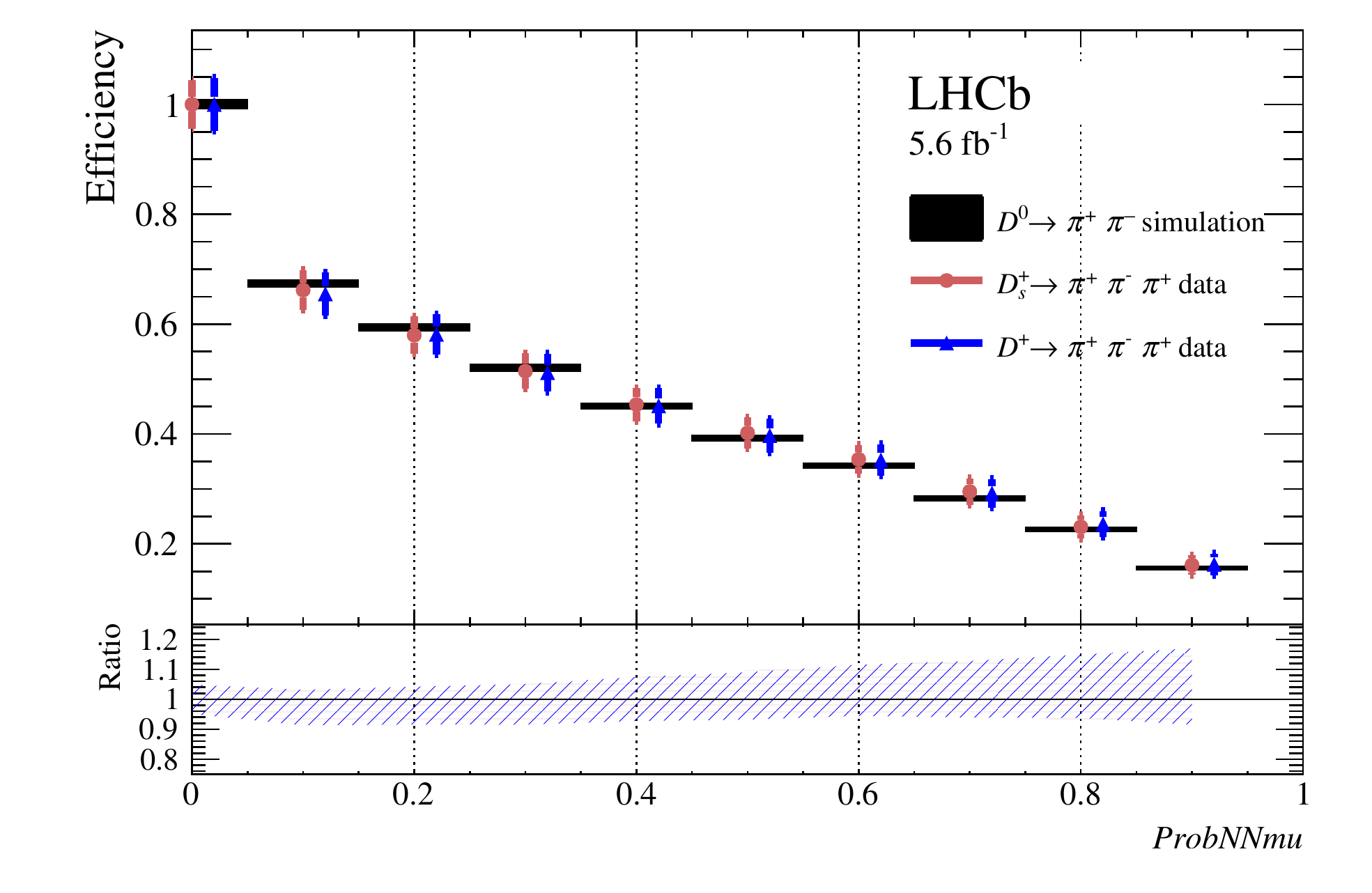}
\caption{Efficiency for two pions to pass a requirement on the $ProbNNmu$ variable (after additional requirements on its identification as muon) in Run 2, as measured from \dpipipi and \dspipipi decays in data and \dpipi decays in simulation. Same sign pions are used in data to avoid contamination for hadronic resonances decaying to two real muons.}\label{fig:pidcalib}
\end{figure}

\begin{figure}
\begin{center}
\includegraphics[width = 0.7\textwidth]{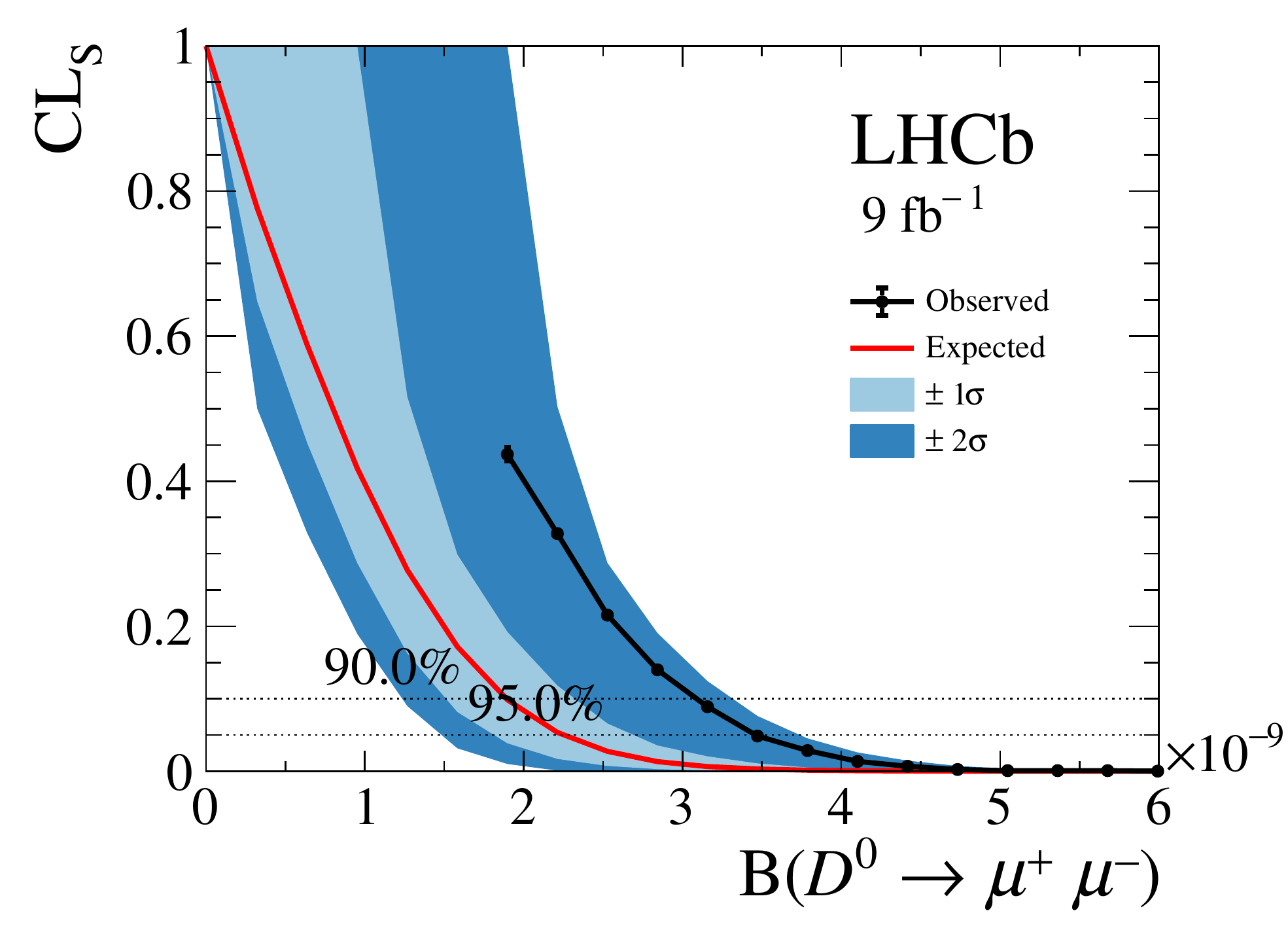}
\end{center}
\caption{Value of $\rm{CL_s}$ as a function of the \dmumu branching fraction. }
\label{fig:cls}
\end{figure}

This section contains the additional figures mentioned in the main text. 
Figures~\ref{fig:signalplot_run1} and \ref{fig:signalplot_run2} present the same data shown in Figure 1 in the main body, but for all the BDT intervals. 
Figures~\ref{fig:signalplot_uncut_run1} and \ref{fig:signalplot_uncut_run2} represent the same distributions and subdivisions of the previous plots, 
but without restricting each to the signal region of the other variable, \ie they contain the full data used for the signal search. 
Figure~\ref{fig:2dplot} shows the data in the two-dimensional plane of the \mmumu and \deltam variables, as well as the signal regions in each variable. 
Figure~\ref{fig:bdtcalib} displays the result of the calibration of the BDT output described in the text. 
Figure \ref{fig:pidcalib} shows the test on the particle identification variable mentioned in the main body. 
Finally, Figure~\ref{fig:cls} shows the value of the $\rm{CL_s}$ estimator used to compute the upper limit on the \dmumu branching fraction, as a function of the branching fraction itself.

\end{document}